\documentclass[pra,twocolumn,showpacs]{revtex4}
\usepackage{graphicx,amsfonts,amsmath,amssymb}
\usepackage{txfonts}

\begin{document}
\bibliographystyle{h-physrev3}
\title{Bright Solitary-Matter-Wave Collisions in a Harmonic Trap: Regimes of Soliton-like Behaviour}
\author{A. D. Martin}
\affiliation{Department of Physics, Durham University, Durham DH1 3LE, United Kingdom}
\author{C. S. Adams}
\affiliation{Department of Physics, Durham University, Durham DH1 3LE, United Kingdom}
\author{S. A. Gardiner}
\affiliation{Department of Physics, Durham University, Durham DH1 3LE, United Kingdom}

\date{\today}
\begin{abstract}
Systems of solitary-waves in the 1D Gross-Pitaevskii equation, which models a trapped atomic Bose-Einstein condensate, are investigated theoretically. To analyse the soliton-like nature of these solitary-waves, a particle analogy for the solitary-waves is formulated.  Exact soliton solutions exist in the absence of an external trapping potential, which behave in a particle-like manner, and we find the particle analogy we employ to be a good model also when a harmonic trapping potential is present.  In the case of two solitons, the particle model is integrable, and the dynamics are completely regular. The extension to three particles supports chaotic regimes. The agreement between the particle model and the wave dynamics remains good even in chaotic regimes. In the case of a system of two solitary waves of equal norm, the solitons are shown to retain their phase difference for repeated collisions. This implies that soliton-like regimes may be found in 3D geometries where solitary waves can be made to repeatedly collide out of phase, stabilising the condensate against collapse.
\end{abstract}

 \maketitle 
\section{Introduction \label{Sec:Introduction}}
Solitary-waves may be found in solutions to nonlinear wave equations where the nonlinearity counteracts the dispersion of a wave-packet such that it retains its form as it propagates. Solitons are solitary-waves that emerge unscathed from collisions with each other, up to shifts in position and phase; this behaviour is reminiscent of particle behaviour, motivating the particle-like name soliton. This distinction is an important one, although in practice the names soliton and solitary-wave are commonly interchanged. ``Classic'' solitons, in this sense, are to be found in integrable nonlinear wave-equations, such as the Korteweg-de Vries equation, the sine-Gordon equation, and the one-dimensional nonlinear Schr\"{o}dinger equation. The solitons' ability to re-emerge after collisions is due to the fact that their dynamics are strongly constrained by conservation laws associated with the wave-equations' integrability \cite{Faddeev_Book_1987}. 

Solitons and solitary-waves are topics of keen interest in the atomic Bose-Einstein condensate (BEC) community.   This is because low-temperature  BEC dynamics are frequently described to a good approximation by the Gross-Pitaevskii equation (GPE) \cite{Dalfovo_RMP_1999,Pethick_Book_2002,Pitaevskii_Book_2003}, a 3D nonlinear wave equation. For regimes where the atoms are confined in the radial direction by a tight trapping potential, the 3D GPE reduces approximately to a 1D equation (the so-called 1D GPE). The homogeneous 1D GPE is simply the 1D nonlinear Schr\"{o}dinger equation, which can be solved by the inverse scattering transform, and yields bright soliton solutions when the nonlinearity is attractive \cite{Zakharov_ZETF_1971,Faddeev_Book_1987}.  At sufficiently low temperatures the interatomic scattering can be largely described through a single parameter, the $s$-wave scattering length.  In this context,  an attractive nonlinearity arises from a negative $s$-wave scattering length, which may be intrinsic, or which may be induced by exploiting a Feshbach resonance to tune the inter-atomic interactions \cite{Inouye_Nature_1998,Kohler_RMP_2006}. As well as describing BEC under tight transverse confinement, the 1D nonlinear Schr\"odinger equation is also used to describe nonlinear optical systems \cite{Haus_RMP_1996,Stegeman_Science_1999}. These systems provide a useful analogue of BEC under tight transverse confinement, and we will frequently refer to work on nonlinear optics in this paper.

Experiments involving BECs composed of attractively interacting atoms have been performed in 1D geometries, resulting in the observation of  single \cite{Khaykovich_Science_2002} and multiple bright solitary-waves \cite{Cornish_PRL_2006,Strecker_AdvSpaceRes_2005,Strecker_Nature_2002}. In the experiments with multiple solitary-waves, the BEC was trapped in the axial direction by a (relatively weak) harmonic confining potential in addition to the radial confinement. The addition of an axial harmonic potential acts to break the integrability of the 1D GPE, meaning that we no longer have exact soliton solutions. In the experiment by Strecker \textit{et al.} \cite{Strecker_AdvSpaceRes_2005,Strecker_Nature_2002}, classic soliton-like behaviour (where the solitary-waves collide and reform up to shifts in phase and position) was not observed, but rather, trains of solitary-waves which are continuously repelled by each other. The dynamics of solitary-wave trains both in BEC and nonlinear optics have been the topic of extensive modeling using a variational method \cite{Al_Khawaja_PRL_2002}, numerical simulations \cite{Carr_PRL_2004,Gawryluk_JPhysB_2006,Hasegawa_OptLett_1984,Konotop_PRA_2002,Leung_PRA_2002,Salasnich_PRL_2003}, a Toda lattice approach, \cite{Gerdjikov_PRE_2006}, a particle model \cite{Gordon_OptLett_1983} (quite distinct to that presented in this paper), analysis using the inverse-scattering transform \cite{Panoiu_PRE_1999} and by using a perturbation approach \cite{Karpman_Physica3D_1981,Kivshar_RMP_1989,Okamawari_PRA_1995}. These treatments model regimes where the solitary-waves are never well separated, where it has been found that the solitary-waves do not collide with each other and re-form, but interact with each other by attractive and repulsive forces, depending on their relative phase.  Motivated by the observation of such soliton trains, a ``soliton laser'' has been proposed  \cite{Chen_JPhysB_2005}. A review article on BEC solitons addresses some of this work in more detail \cite{Abdullaev_IntJModPhysB_2005}.

As opposed to solitary-wave trains, we investigate whether classic soliton-like behaviour, i.e., colliding and reforming of distinct, localized wave packets up to shifts in phase and position, is possible in the 1D GPE with a harmonic potential. In a previous work \cite{Martin_PRL_2007} we found regimes where such behaviour is quite pronounced. This behaviour was also seen in work done in similar nonlinear optical settings \cite{Scharf_CSF_1995,Scharf_PRE_1993,Scharf_PRA_1992,Elyutin_PRE_2001}. In this paper we further our investigation into soliton-like behaviour; in particular we explore the bounds within which the solitary-waves can still be expected to behave as solitons. To this end, we use a particle model introduced in our previous work \cite{Martin_PRL_2007}, adapted from a model developed for use in nonlinear optics \cite{Scharf_CSF_1995,Scharf_PRE_1993,Scharf_PRA_1992}. We show that soliton-like behaviour is possible in the 1D GPE with a harmonic potential, provided that the solitary-waves collide with large enough relative velocity such that the collisions occur during a short timescale compared with the period of the axial trapping potential. This type of behaviour has recently been experimentally observed \cite{Cornish_PRL_2006}, and provides an exciting prospect for future experiments to probe the dynamics in more detail.

In the case of three solitons, we find regimes of regular and chaotic dynamics.  In particular, chaotic solutions to the GPE are expected to coincide with more rapid condensate depletion than in otherwise similar regular solutions \cite{Gardiner_JMO_2002}; indeed this has been seen in theoretical studies of several systems \cite{Castin_PRL_1997,Gardiner_PRA_2000,Zhang_PRL_2004,Liu_PRA_2006,Reslen_unpublished_2007}. This provides an additional motivation to identify regimes of regular and chaotic soliton dynamics in the GPE.

In more realistic models for BECs, the integrability of the nonlinear wave equation is also broken by residual 3D effects. These effects cause the soliton collisions to be inelastic; specifically, there is particle exchange between the solitons accompanied by changes in their outgoing velocities \cite{Parker_unpublished_2006}. A reduction from 3D to non-integrable 1D equations
\cite{Kamchatnov_PRA_2004,Salasnich_PRA_2002}, more sophisticated than the 1D GPE, confirms this result \cite{Khaykovich_PRA_2006}. This type of behaviour is common in other non-integrable Schr\"odinger-type equations: \cite{Dmitriev_PRE_66_2002,Dmitriev_PRE_68_2003,Goodman_PRL_2007,Papacharalampous_PRE_2003}. Bose-Einstein condensates with attractive interatomic interactions are prone to collapse if the particle density becomes too high \cite{Cornish_PRL_2006}. Fully 3D GPE simulations show that in-phase collisions between solitons, during which the particle density becomes large, can cause collapse of the condensate, \cite{Parker_unpublished_2006,Cornish_PRL_2006,Parker_JPhysB_2007}. In this paper, as well as identifying regimes in which soliton-like behaviour can still occur despite the mild breaking of integrability by the harmonic potential, we also briefly discuss regimes where solitons are expected to survive 3D integrability breaking.

The layout of the paper is as follows: in section II we introduce the model equations and reiterate the soliton solution to the homogeneous 1D GPE; in section III we identify one, two and three solitary-wave solutions to the 1D GPE with a harmonic potential and introduce a particle model to test their soliton nature; in section IV we present our conclusions.

\section{Background \label{Sec:Background}}

\subsection{Model system \label{Sec:ModelSystem}}

\subsubsection{Quantum field \label{Sec:QuantumField}}

In the case of an atomic Bose gas, the dynamics of the system may be described, in
the Heisenberg picture, by the time-evolution of the bosonic field operator
$\hat{\Psi}(\mathbf{r})$. At the temperatures typically encountered in atomic BEC experiments (nK), for sufficiently dilute Bose gases, atomic interactions are dominated by low energy two-body collisions. In this case, the atom-atom interactions are characterised by one parameter: the $s$-wave scattering length, $a$. Moreover, we may generally replace the true interaction potential by an effective contact interaction, subject to an appropriate renormalization procedure \cite{Stoof_PRE_1993,Rusch_PRA_1999,Morgan_JPB_2000}.  Hence, we let $V_{{\mbox{\scriptsize
int}}}(\mathbf{r},\mathbf{r}^{\prime})=g_{\mbox{\scriptsize 3D}}\delta(\mathbf{r}-\mathbf{r}^{\prime})$ where $g_{\mbox{\scriptsize 3D}}=4\pi\hbar^{2}a /m$ and $m$ is the particle mass of the species. Depending on the species, $a$ may be positive or negative, corresponding to an effective repulsive or attractive interaction. By exploiting a Feshbach resonance, it may also be tuned using an external magnetic field \cite{Inouye_Nature_1998,Kohler_RMP_2006}. 

The many-body Hamiltonian for the system in second-quantized form is then given by:
\begin{equation}
\hat{H}=\int
d\mathbf{r}\hat{\Psi}^{\dag}(\mathbf{r})\left[\mathcal{H}+\frac{g_{\mbox{\scriptsize 3D}}}{2}\hat{\Psi}^{\dag}(\mathbf{r})\hat{\Psi}(\mathbf{r})
\right] \hat{\Psi}(\mathbf{r}),
\end{equation}
where 
\begin{equation}
\mathcal{H} =
-\frac{\hbar^{2}\nabla^{2}}{2m}+V_{\mbox{\scriptsize
ext}}(\mathbf{r})
\end{equation}
is the Hamiltonian for a single particle in
the external trapping potential. If Bose-Einstein condensation has occurred, we may define the condensate mode as the eigenfunction $\Psi(\mathbf{r})$ of the single-body density matrix $\rho(\mathbf{r},\mathbf{r}^{\prime})=\langle \hat{\Psi}^{\dagger}(\mathbf{r}^{\prime})\hat{\Psi}(\mathbf{r})\rangle$ with the largest eigenvalue \cite{Penrose_PR_1956,Castin_PRA_1998,Gardiner_PRA_2007}. We are then free to partition the field operator into condensate and non-condensate parts \cite{Gardiner_PRA_1997,Castin_PRA_1998,Gardiner_PRA_2007}:
\begin{equation}
\hat{\Psi}(\mathbf{r})=\hat{a}\Psi(\mathbf{r})+\delta \hat{\Psi}(\mathbf{r}),
\end{equation}
where $\hat{a}$ annihilates a particle in mode $\Psi(\mathbf{r})$, and $\delta\hat{\Psi}(\mathbf{r})$ is the field operator for modes orthogonal to the condensate.

\subsubsection{3D classical field}
In the case of a trapped, almost fully Bose-condensed dilute atomic gas, the dynamics of the condensate mode $\Psi(\mathbf{r})$ are largely governed by the following GPE \cite{Castin_PRA_1998,Gardiner_PRA_2007}:
\begin{equation}
i \hbar \frac{\partial}{\partial t} \Psi(\mathbf{r},t)=\left[
-\frac{\hbar^{2}\nabla^{2}}{2m}+V_{\mbox{\scriptsize
ext}}(\mathbf{r})+g_{\mbox{\scriptsize 3D}}N\left|\Psi(\mathbf{r},t) \right|^{2}
\right]\Psi(\mathbf{r}),\label{GP}
\end{equation}
where $N$ is the total number of particles in the condensate and
$\Psi(\mathbf{r})$ is normalised to one.

We now consider a cylindrically symmetric (cigar-shaped) harmonic trapping potential:
\begin{equation}
V_{\mbox{\scriptsize
ext}}(\mathbf{r})=\frac{m}{2}\left[\omega_{x}^{2}x^{2}+\omega_{r}^{2}(y^{2}+z^{2})\right],
\end{equation}
where $\omega_{x}\ll\omega_{r}$. We also explicitly assume $a<0$ (attractive inter-particle interactions), and determine that Eq.\ (\ref{GP}) reduces to the following 1D equation (see appendix \ref{App:3Dto1D} for a derivation):
\begin{equation}
i\frac{\partial}{\partial t}\psi(x)=-\frac{1}{2}\frac{\partial^{2}}{\partial x^{2}}\psi(x)+\frac{\omega^{2}x^{2}}{2}\psi(x)-
|\psi(x)|^{2}\psi(x) \label{S2},
\end{equation}
where $x$ is measured in units of
$\hbar^{2}/m|g_{\mbox{\scriptsize 1D}}|N$ and
$t$ in units of
$\hbar^{3}/m|g_{\mbox{\scriptsize 1D}}|^{2}N^{2}$,
with $g_{\mbox{\scriptsize 1D}}=2\hbar \omega_{r}a$ and $\omega$ equal to the axial frequency $\omega_{x}$ in our units of inverse time ($m|g_{\mbox{\scriptsize 1D}}|^{2}N^{2}/\hbar^{3}$).

\subsubsection{Linear instability and chaos \label{Sec:LinearInstability}}
It can be shown that linear instabilities in the GPE directly imply, in both Bogoliubov \cite{Bogoliubov_JPhys_1947} and equivalent number-conserving linearised approaches \cite{Castin_PRA_1998}, that the population of the non-condensate component may rapidly become significant. For this reason we expect regimes where the GPE dynamics are chaotic to coincide with rapid depletion of the condensate. The GPE is a norm-conserving equation, so the depletion will not show up in the GPE dynamics; however, we may use chaos in GPE dynamics as an indicator of depletion of the condensate mode in a realistic system. This motivates the identification of chaotic trajectories in the GPE dynamics, as we discussed in the introduction (Sec.\ \ref{Sec:Introduction}). 

\subsection{Soliton solution to homogeneous GPE}
In the case of no axial potential [equivalent to setting $\omega=0$ in Eq.\ (\ref{S2})], it is possible to find exact solutions of Eq.\ (\ref{S2}) \cite{Zakharov_ZETF_1971}.  A straightforward interpretation is in terms of a scattering problem \cite{Faddeev_Book_1987}; in the limit $t \rightarrow - \infty$, the solutions take the form of an arbitrary number of well separated (incoming) solitons:
\begin{equation} 
\Phi_{j}(x,t)=2\eta_{j} \mathrm{sech} \left[2\eta_{j}(x-q_{j})\right]
e^{i v_{j}(x-q_{j})}e^{i(2\eta_{j}^{2}+v_{j}^{2}/2) t}e^{i\alpha_{0j}}.
\label{Sol_1}
\end{equation}
Here $q_{j}=v_{j}t +x_{0j}$ is the position of the peak of the $j$th soliton; $x_{0j}$ is the peak position at $t=0$; $\alpha_{0j}-v_{j}x_{0j}$ is the phase for a single soliton (i.e., in the absence of collisions with other solitons) at $x=0$, $t=0$; $v_{j}$ is the soliton velocity and $\eta_{j}$ gives the relative size of the soliton. Our normalisation condition implies $\sum_{j}^{N_{s}}4\eta_{j}=1$, where $N_{s}$ is the number of solitons present.

The solitons come together and collide, during which time the form of the solution is complicated and solitons are not individually defined. However, as $t\rightarrow \infty$, the outgoing solitons re-emerge from the collisions unscathed, taking the same asymptotic form [Eq.\ (\ref{Sol_1})], up to shifts in position and phase: $q_{j}\mapsto q_{j}+\delta x_{j}$ and $\alpha_{0j}\mapsto\alpha_{0j}+\delta \phi_{j}$, where the position shift $\delta x_{j}$ and phase-shift $\delta\phi_{j}$ of the $j$th soliton are given by \cite{Zakharov_ZETF_1971,Faddeev_Book_1987}:
\begin{equation}
2\eta\delta x_{j}+i\delta\phi_{j}=\sum_{j\neq k}\pm 2  \ln\left[ \frac{v_{j}-v_{k}+i2(\eta_{j}+\eta_{k})}{v_{j}-v_{k}+i2(\eta_{j}-\eta_{k})}\right]. \label{Shifts}
\end{equation}
The positive sign applies if the soliton is on the left prior to the collision with the $k$th soliton ($v_{j}>v_{k}$), otherwise the negative sign applies. Note that these shifts are dependent on the solitons' initial speeds $v_{j}$, and effective masses $\eta_{j}$ only, not on their relative phase. 

\section{Soliton dynamics with a harmonic external potential \label{Sec:SolitonDynamics}}

\subsection{Single soliton \label{Sec:SingleSoliton}}

\begin{figure}[tbp]
\centering
\includegraphics[width=8.8cm]{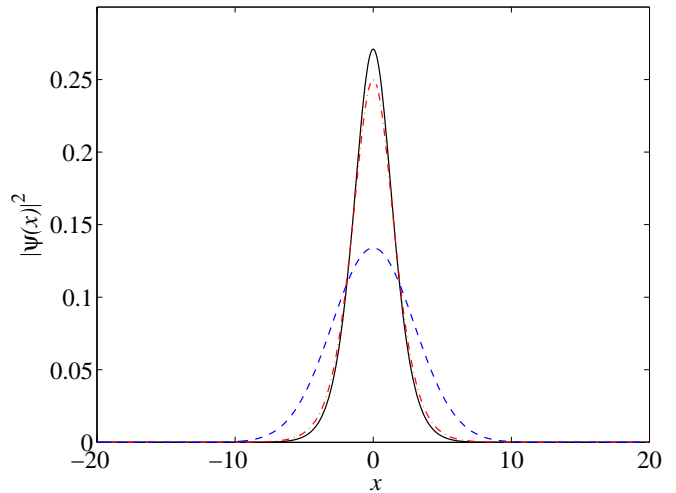}
\caption{(Colour online). Ground state solution for harmonically trapped soliton of 5000 particles (solid line). Corresponding soliton solution to the homogeneous equation (dot-dashed line), which is used as an ansatz in the particle model. The ground state of the linear Schr\"odinger equation (dashed line) is given for comparison. The parameters of the system are taken to be similar to those of a recent experiment \cite{Strecker_Nature_2002} the axial trapping frequency is $10/2\pi$ Hz, the radial trap frequency is $800/2\pi$ Hz, atomic mass and scattering length of $^{7}$Li. A unit of $x$ is hence equal to $ 7.19\times 10^{-6}$ m.} \label{Anz}
\end{figure}

As shown in appendix B, for any solution to the 1D harmonic GPE, there exist other solutions with the same density profile, undergoing arbitrary amplitude harmonic oscillations at the trap frequency.  In particular, for any stationary solution, there exist corresponding solutions with the same density profile, which oscillate with the trap frequency but remain otherwise unchanged.  Hence, a single bright soliton in a harmonic trap experiences an overall simple harmonic motion without any manifestation of internal dynamics in the soliton's density profile. 

The density profile and phase behaviour of a single soliton can be found by first considering the form of a stationary soliton, and then inferring the behaviour of the oscillating version. The stationary soliton will be a solution to the eigenvalue problem:
\begin{equation}
-\frac{1}{2}\frac{\partial^{2}}{\partial x^{2}}\psi(x)+\frac{\omega^{2}x^{2}}{2}\psi(x)-
|\psi(x)|^{2}\psi(x)=\mu\psi(x), \label{TINLSE}
\end{equation}
and will have the form:
$\psi(x)=u(x)\exp\left\{-i\left[\mu t +S(0)\right]\right\}$, where $u(x)$ is a real valued function and the real-valued number $S(0)$ is an initial phase. We expect the single stationary soliton solution to be the metastable ``ground'' state of the system \cite{Ruprecht_PRA_1995,Kagan_PRL_1996,Pitaevskii_PLA_1996}, which may be determined numerically, for example by propagating Eq.\ (\ref{S2}) in imaginary time \cite{Chiofalo_PRE_2000,Lehtovaara_JComputPhys_2007}. The numerically determined density $u(x)^{2}$ for a parameter regime consistent with the $^{7}$Li experiments of Strecker \textit{et al}.\ is shown in figure \ref{Anz}, and is compared to a bright soliton solution of the homogeneous 1D GPE, and to the ground state of the 1D linear Schr\"odinger equation with harmonic potential. As expected, the solution of the 1D GPE with a harmonic potential is spatially slightly compressed compared to the bright soliton solution of the homogeneous GPE.  These two solutions, however, are quite similar (and can be made more similar as $\omega$ is progressively reduced), and are quite distinct from the Gaussian solution produced by the linear Schr\"{o}dinger equation. We will exploit this similarity later in the paper.

As shown by the treatment in appendix \ref{App:HarmonicallyOscillating}, an oscillating soliton solution takes the form:
\begin{equation}
\psi(x,t)=u\left[x-\langle x(t) \rangle\right]\exp\left(i\left[-\mu t+\langle p(t)\rangle x-S(t)\right]\right),\label{coherent_state}
\end{equation}
where $\langle x(t) \rangle=x_{0}\cos(\omega t)+(p_{0}/\omega) \sin(\omega t)$ is the position expectation value of the atomic ensemble, $\langle p(t)\rangle=p_{0}\cos(\omega t)-\omega x_{0}\sin(\omega t)$ is the momentum expectation value, i.e., 
\begin{equation}
\langle x\rangle=\int dx \psi^{*}(x)x\psi(x)
\end{equation}
and  
\begin{equation}
\langle p\rangle= -i\int dx \psi^{*}(x)
\frac{\partial \psi(x)}{\partial x};
\end {equation}
$p_{0}$ and $x_{0}$ are the initial position and momentum expectation values, respectively, and 
\begin{equation}
S(t)=\left(p_{0}^{2}-\frac{x_{0}^{2}\omega^{2}}{2}\right)\frac{\sin(2\omega t)}{2\omega}+\frac{x_{0}p_{0}}{2}\cos(2\omega t)-\frac{x_{0}p_{0}}{2} + S(0). \label{Action}
\end{equation}
Removing the nonlinearity reproduces the result for a coherent state. When the nonlinearity is present, however, the stationary eigenvalue, $\mu$, is dependent on the norm of the soliton, unlike the case of the linear Schr\"odinger equation.

\subsection{Two solitons \label{Sec:TwoSolitons}}

\begin{figure*}[tbp]
\centering
\includegraphics[width=17.6cm]{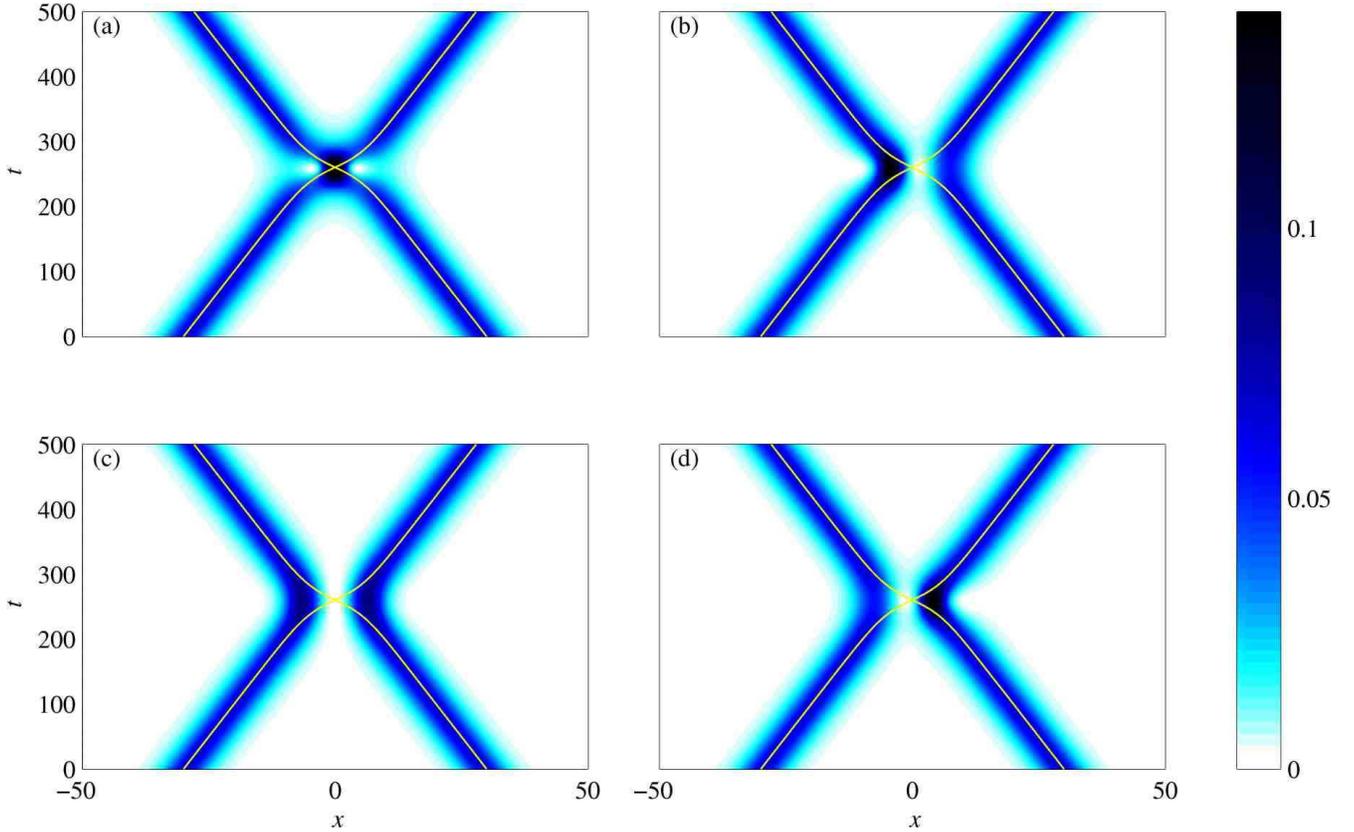}
\caption{(colour online). Two soliton collision taking place when (a) $\Delta\phi_{\mathrm{col}}=0$, (b) $\Delta\phi_{\mathrm{col}}=\pi/2$, (c) $\Delta\phi_{\mathrm{col}}=\pi$ and (d) $\Delta\phi_{\mathrm{col}}=3\pi/2$. The parameters of the system are taken to be similar to those of a recent experiment \cite{Strecker_Nature_2002} the axial trapping frequency is $10/2\pi$ Hz, the radial trap frequency is $800/2\pi$ Hz, atomic mass and scattering length of $^{7}$Li, and 5000 particles per soliton). The unit of $x$ is then equal to 3.6 $\mu$m, and a unit of $t$ to 1.4 ms.} \label{fig_phase1}
\end{figure*}

\subsubsection{Overview \label{Sec:TwoSolitonsOverview}}

\begin{figure*}[tbp]
\centering
\includegraphics[width=17.6cm]{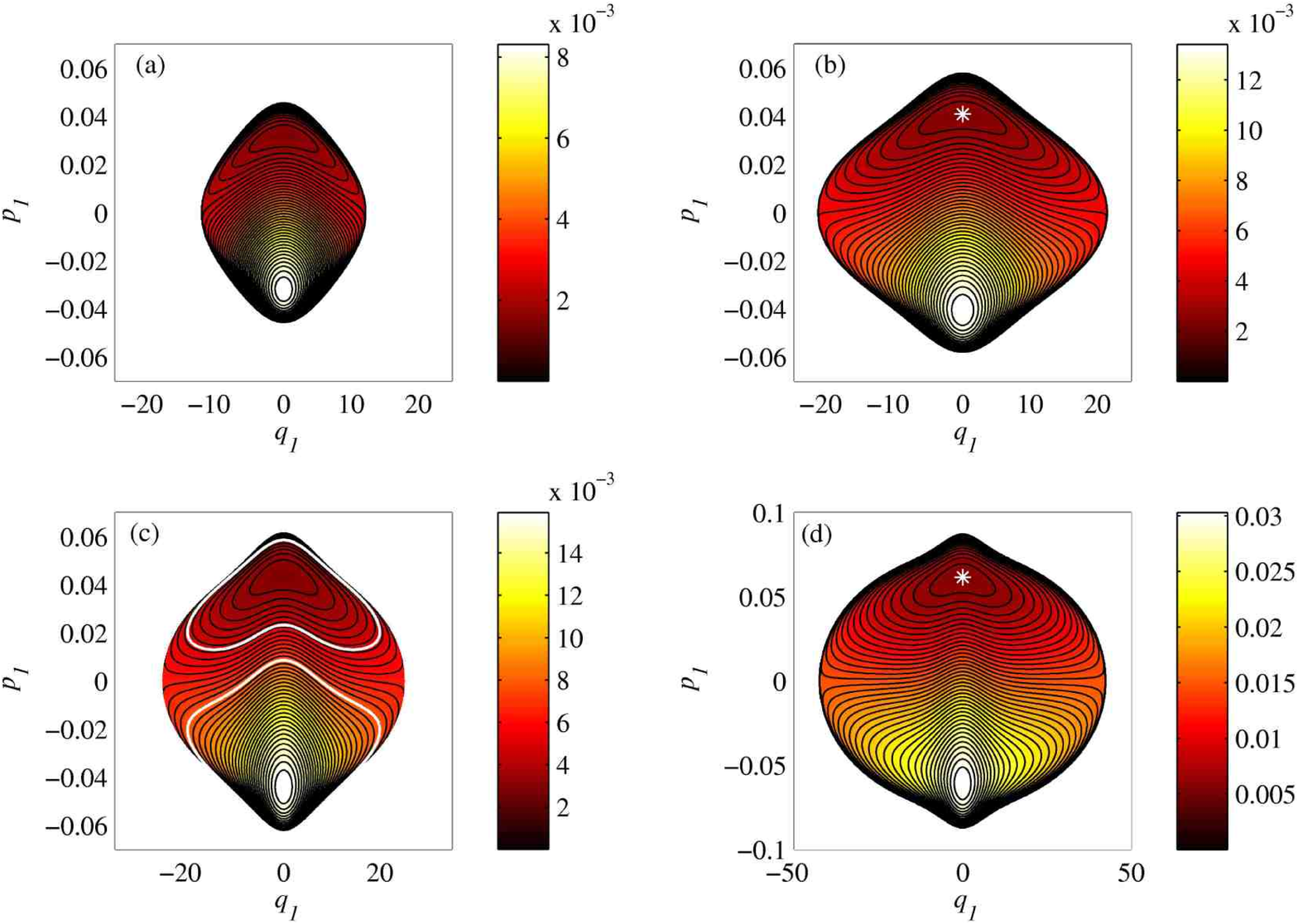}
\caption{(Colour online) Poincar\'e sections for the two-soliton system corresponding to the momentum $p_{1}$ and position $q_{1}$ of one soliton, while the other soliton has coordinates $q_{2}=0$, $p_{2}<0$. The value of centre-of-mass energy, $E$, in the is given by the colour scale. (a) Total energy $H=5\times 10^{-4}$; (b) $H\approx 5.6\times 10^{-3}$, the star corresponds to the trajectory in figure \ref{fig_two_b}; (c) $H\approx 8.1\times 10^{-3}$, the upper trajectory correspond to that in figure \ref{fig_two_c}, the lower to that in figure \ref{fig_two_c2}; (d) $H\approx 2.2\times 10^{-2}$, the star corresponds to the trajectory in figure \ref{fig_two_d}. The figures correspond to regimes where the solitons have equal effective masses, the axial trapping frequency is $10/2\pi$ Hz, and the other parameters (radial trap frequency of $800/2\pi$ Hz, atomic species mass and scattering length of $^{7}$Li, and 5000 particles per soliton) are comparable to those in recent experiment \cite{Strecker_Nature_2002}.} \label{fig_poin_2}
\end{figure*}

In this section, we present a simple model of multiple trapped solitons, treating each of the solitons as a classical particle. We explore the case of two harmonically trapped solitons, and present results comparing the trajectories in the particle model with simulations of the wave dynamics in the GPE. These results allow us to determine the range of initial conditions for which the particle model is a good description of the system. 

\subsubsection{Particle model \label{Sec:ParticleModel}}

Recall from Sec.\ \ref{Sec:Background} that, in a homogeneous system, the trajectories of solitons emerging from collisions with each other are independent of the relative phase of the incoming solitons. The only effect of the relative phase of the solitons is on the form of the wavefunction (peak or trough) during the collision. This is illustrated in figure \ref{fig_phase1}. The phase-independence of the solitons' incoming and outgoing trajectories allows a model to be formulated that treats the solitons as classical particles, each with only the positional degree of freedom (rather than position and phase degrees of freedom used, for example, in \cite{Gordon_OptLett_1983}). This model was introduced by Scharf and Bishop in the context of nonlinear optics \cite{Scharf_CSF_1995,Scharf_PRE_1993,Scharf_PRA_1992}, which we have adapted for the purpose of modeling a quasi-1D harmonically trapped BEC \cite{Martin_PRL_2007}.

To construct the particle model, we first consider the homogeneous solution before introducing the effects of the harmonic trap. Following the approach in \cite{Maki_PRL_1986}, one can derive  an effective inter-soliton potential (see appendix \ref{App:Deduction}):
\begin{equation}
V(q_{j}-q_{k})=-2\eta_{j}\eta_{k}(\eta_{j}+\eta_{k})\mathrm{sech}^{2}\left[\frac{2\eta_{j}\eta_{k}}{\eta_{j}+\eta_{k}}(q_{j}-q_{k}) \right], \label{V_interaction}
\end{equation} 
which treats the solitons as particles of position $q_{j}$ and effective mass $\eta_{j}$, the parameters used to describe the bright soliton solutions of Eq.\ (\ref{Sol_1}).  This potential reproduces the asymptotic position shifts [Eq.\ (\ref{Shifts})] in the homogeneous GPE for the outgoing particle trajectories, i.e., the position shifts as the solitons become infinitely far apart. It yields accurate results when $2|\eta_{1}-\eta_{2}|\ll |v_{1}-v_{2}|$, where $v_{1}$ and $v_{2}$ are the soliton velocities, and which gives a lower limit for the relative velocity for which the particle model is applicable. 

Figure \ref{fig_phase1} shows the particle trajectories predicted by our model interaction potential [Eq.\ \ref{V_interaction}] superimposed on the density profile dynamics predicted by solution of the homogeneous 1D GPE. When modelling BEC dynamics, an upper limit to the solitons' relative velocity is also imposed, because the contact-interaction potential between atoms, used to derive the GPE, assumes low energy inter-atomic collisions, and may not be applicable to condensates with high relative approach speeds \cite{Kohler_PRA_2002}. Fortunately, recent experiments \cite{Strecker_Nature_2002,Cornish_PRL_2006} show that solitons are generated with similar sizes, such that their velocities may easily fall within our model's range of validity.

In Sec.\ \ref{Sec:SingleSoliton} we showed that independent trapped solitons oscillate harmonically. A more general method for deriving the approximate motion of solitons in an external potential of arbitrary form was found by Scharf and Bishop \cite{Scharf_PRA_1992}, and is outlined in appendix \ref{App:Motion} for completeness. To combine the effects of the external potential and the soliton collisions, we use the homogeneous solution [Eq.\ (\ref{Sol_1})] as an ansatz, so that the solitons are still characterised by the parameters $q_{j}$ and $\eta_{j}$.  Figure \ref{Anz} shows this to be a reasonable approximation.  The following Hamiltonian:
\begin{equation}
\begin{split}
H= & \sum^{N_{s}}_{j=1}\left(\frac{p_{j}^{2}}{2\eta_{j}} + \frac{\eta_{j} \omega^{2} q^{2}_{j}}{2}\right)  
\end{split}
\end{equation} reproduces the harmonic motion of the solitons, keeping the interpretation of $\eta_{j}$ as effective masses (see appendix \ref{App:Motion}). We assume that the soliton-soliton interactions are not affected by the introduction of the (relatively loose) harmonic trap and construct the full Hamiltonian by adding in the contributions from the interaction potentials:
\begin{equation}
\begin{split}
H= & \sum^{N_{s}}_{j=1}\left(\frac{p_{j}^{2}}{2\eta_{j}} + \frac{\eta_{j} \omega^{2} q^{2}_{j}}{2}\right)
\\&-\sum_{1\leq j< k\leq
N_{s}} 2\eta_{j}\eta_{k}(\eta_{j}+\eta_{k})\mathrm{sech}^{2}\left[\frac{2\eta_{j}\eta_{k}}{\eta_{j}+\eta_{k}}(q_{j}-q_{k}) \right],  \label{Ham4}
\end{split}
\end{equation} where $N_{s}$ is the number of solitons. This approach is expected to be valid for regimes when the timescale of the soliton-soliton collisions is much less than the period of the harmonic trap, such that the effects of the harmonic trap are negligible during the collisions. The limits of this approach are further explored in Secs.\ \ref{Sec:Presentation} and \ref{Sec:Discussion}.

In the case of two solitons ($N_{s}=2$), it is useful to define the following independent coordinates: the centre-of-mass position $Q:=(\eta_{1}q_{1}+\eta_{2}q_{2})/(\eta_{1}+\eta_{2}) $ and the relative position $q:=q_{1}-q_{2}$. The Hamiltonian [Eq. (\ref{Ham4})] then takes the form:
\begin{equation}
\begin{split}
 H=&\frac{P^{2}}{2(
\eta_{1}+\eta_{2})} +\frac{\omega^{2}}{2}(\eta_{1}+\eta_{2})Q^{2} 
\\&+\frac{\eta_{1}+\eta_{2}}{2\eta_{1}\eta_{2}}p^{2}+\frac{\omega^{2}}{2}\frac{\eta_{1}\eta_{2}}{\eta_{1}+\eta_{2}}q^{2}
\\&-2\eta_{1}\eta_{2}(\eta_{1}+\eta_{2})\mathrm{sech}^{2}\left(\frac{2\eta_{1}\eta_{2}}{\eta_{1}+\eta_{2}} q\right),\label{Ham_2S}
\end{split}
\end{equation} 
where $P=p_{1}+p_{2}$ is the momentum canonically conjugate to $Q$, and $p=(\eta_{2}p_{1}-\eta_{1}p_{2})/(\eta_{1}+\eta_{2})$ the momentum conjugate to $q$. 
The Hamiltonian is now clearly seperable into two parts: the centre-of-mass energy $E$ (dependent on $P$ and $Q$ only), and the interaction energy $\epsilon$ (dependent on $p$ and $q$ only). There are thus two independent constants of the motion, $E$ and $\epsilon$, as many as there are degrees of freedom. Hence, the particle model for two solitons is integrable and the dynamics must be completely regular \cite{Gutzwiller_Book_1990,Reichl_Book_1992}. 
In the case where the solitons have identical effective masses ($\eta_{1}=\eta_{2}:=\eta$), the Hamiltonian [Eq. (\ref{Ham_2S})] reduces to 
\begin{equation}
\begin{split}
 H=&\frac{P^{2}}{4
\eta} +\eta \omega^{2}Q^{2} +\frac{p^{2}}{\eta}+\frac{ \eta\omega^{2}q^{2}}{4}-4\eta^{3}\mathrm{sech}^{2}\left(\eta q\right).
\end{split}
\label{Ham_2S_identical}
\end{equation} 

Figure \ref{fig_poin_2} shows four Poincar\'e surfaces of section (or Poincar\'e sections) for the two particle system; sections of $p_{1}$ versus $q_{1}$ are shown for different surfaces of different total energy \cite{Gutzwiller_Book_1990,Reichl_Book_1992}. These Poincar\'e sections demonstrate the regular behaviour of the integrable two particle system, as all trajectories lie on invariant tori in the phase space of the system. There are two distinct regimes observable in these Poincar\'e sections. In the lower regions of the sections, the centre-of-mass energy, $E$, is large and positive; in this case the interaction energy, $\epsilon$, has a large negative contribution from the interaction term, and the solitons interact strongly. It is seen in section \ref{Sec:Presentation} that in this regime, there is rapid energy exchange between the solitons, such that the soliton with lower amplitude oscillations is driven by the other soliton, which itself becomes damped. In the upper regions of the sections, $E$ is less positive, and hence $\epsilon$ is less negative, so the energy exchange between the solitons occurs over a longer time period. 

\subsubsection{Presentation of results comparing GPE to particle evolutions \label{Sec:Presentation}}

\begin{figure*}[tbp]
\centering
\includegraphics[height=12cm]{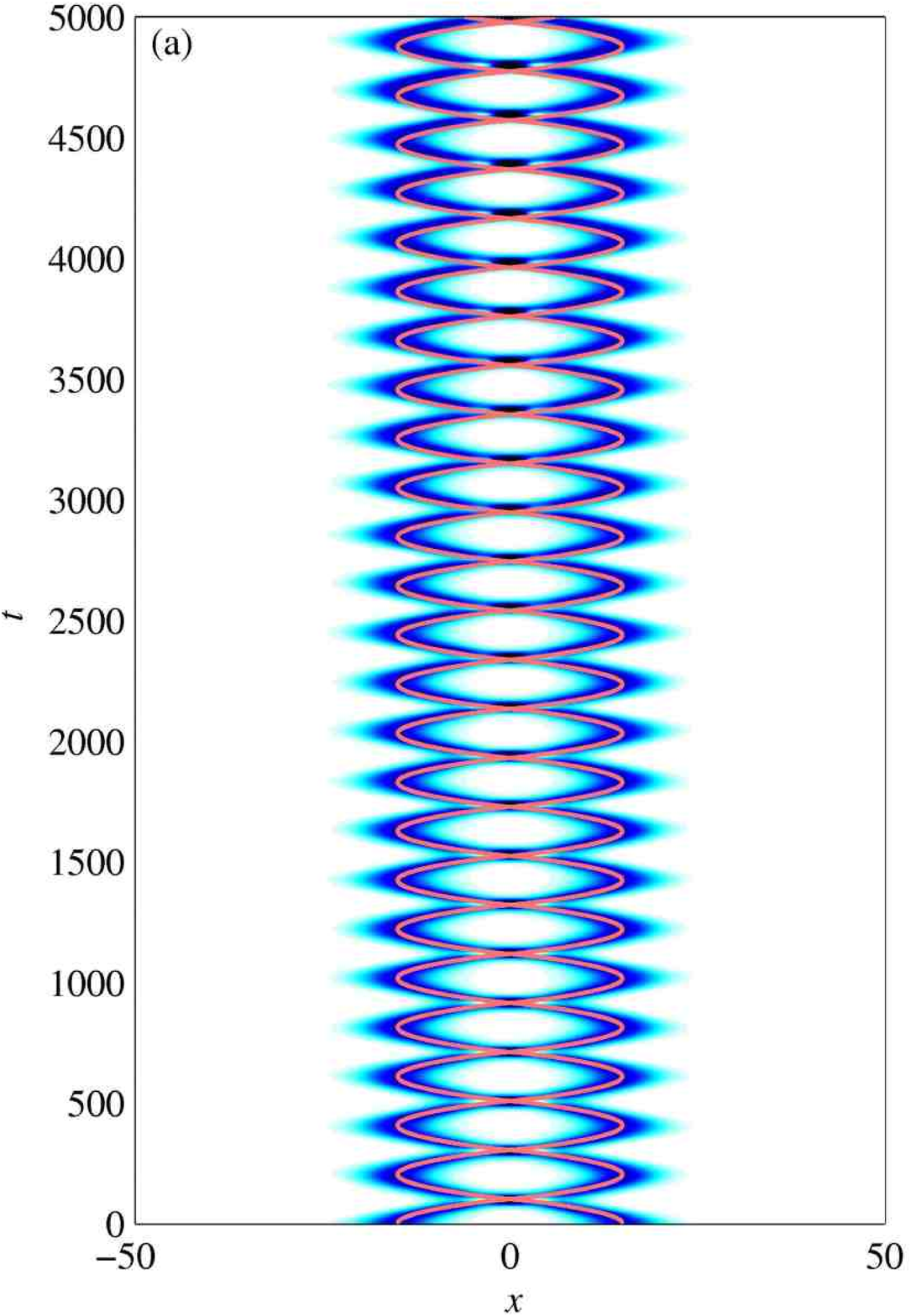}
\includegraphics[height=12cm]{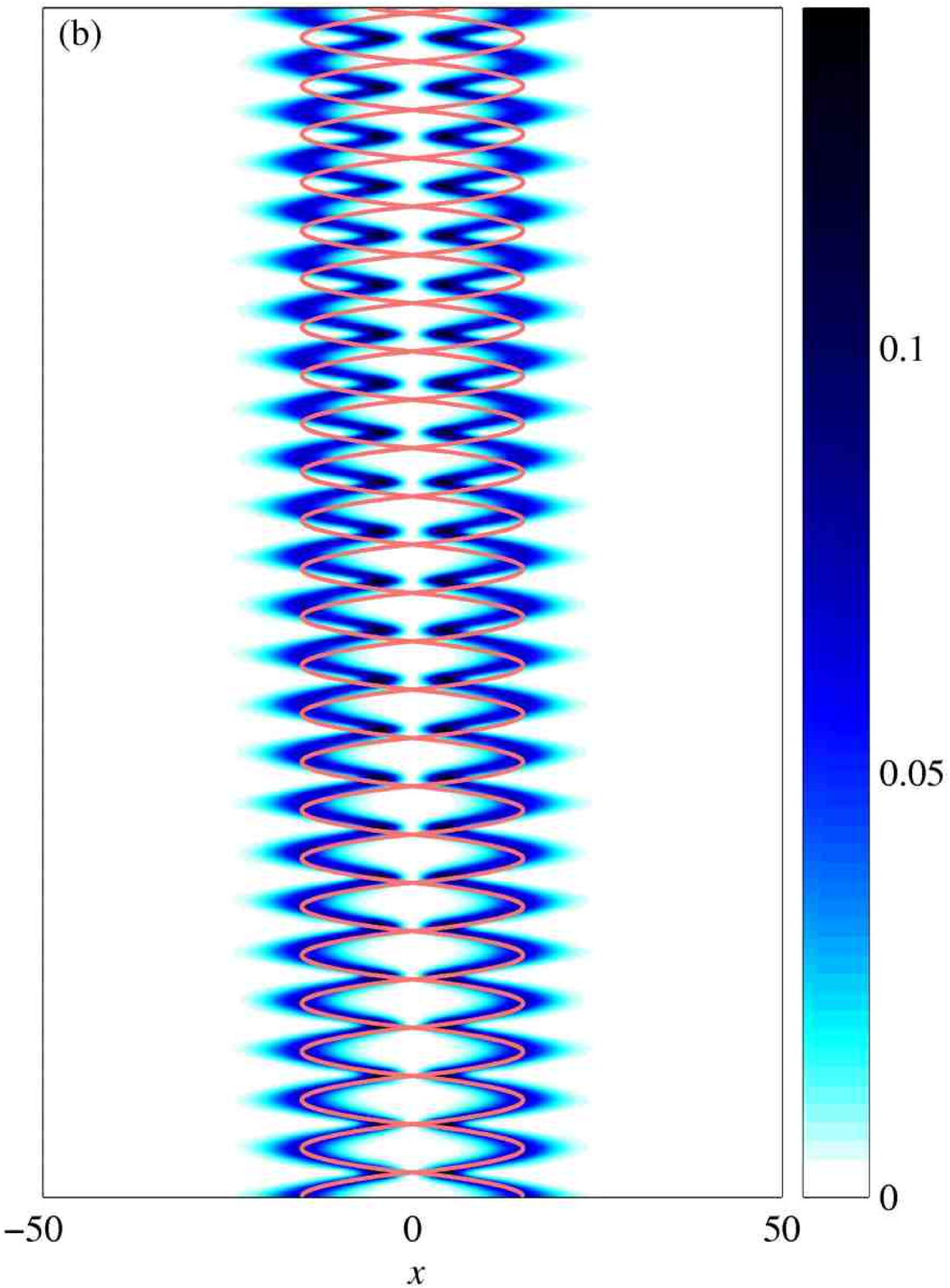}
\caption{(colour online). Trajectories in the particle model (lines) plotted over density distributions predicted by 1D GPE dynamics, corresponding to the trajectory marked on figure \ref{fig_poin_2}(b). The relative phase of the solitons in the wave dynamics is zero in figure (a), and $\pi$ in figure (b). The figures correspond to regimes where the solitons have equal effective masses, the axial trapping frequency is $10/2\pi$ Hz, and the other parameters (radial trap frequency of $800/2\pi$ Hz, atomic species mass and scattering length of $^{7}$Li, and 5000 particles per soliton) are comparable to those in recent experiment \cite{Strecker_Nature_2002}. The unit of $x$ is then equal to 3.6 $\mu$m, and a unit of $t$ to 1.4 ms. } \label{fig_two_b}
\end{figure*}

\begin{figure*}[tbp]
\centering
\includegraphics[height=12cm]{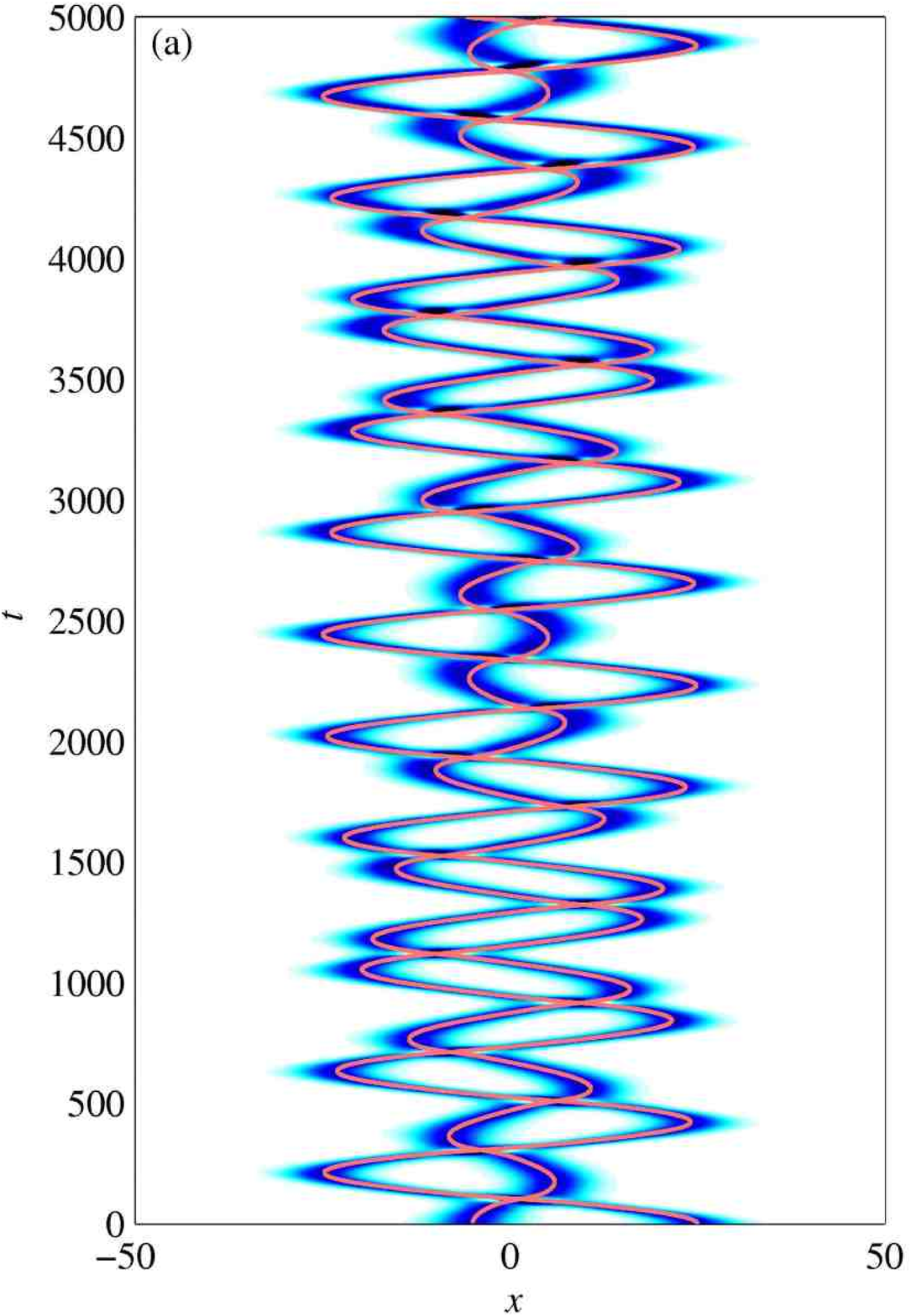}
\includegraphics[height=12cm]{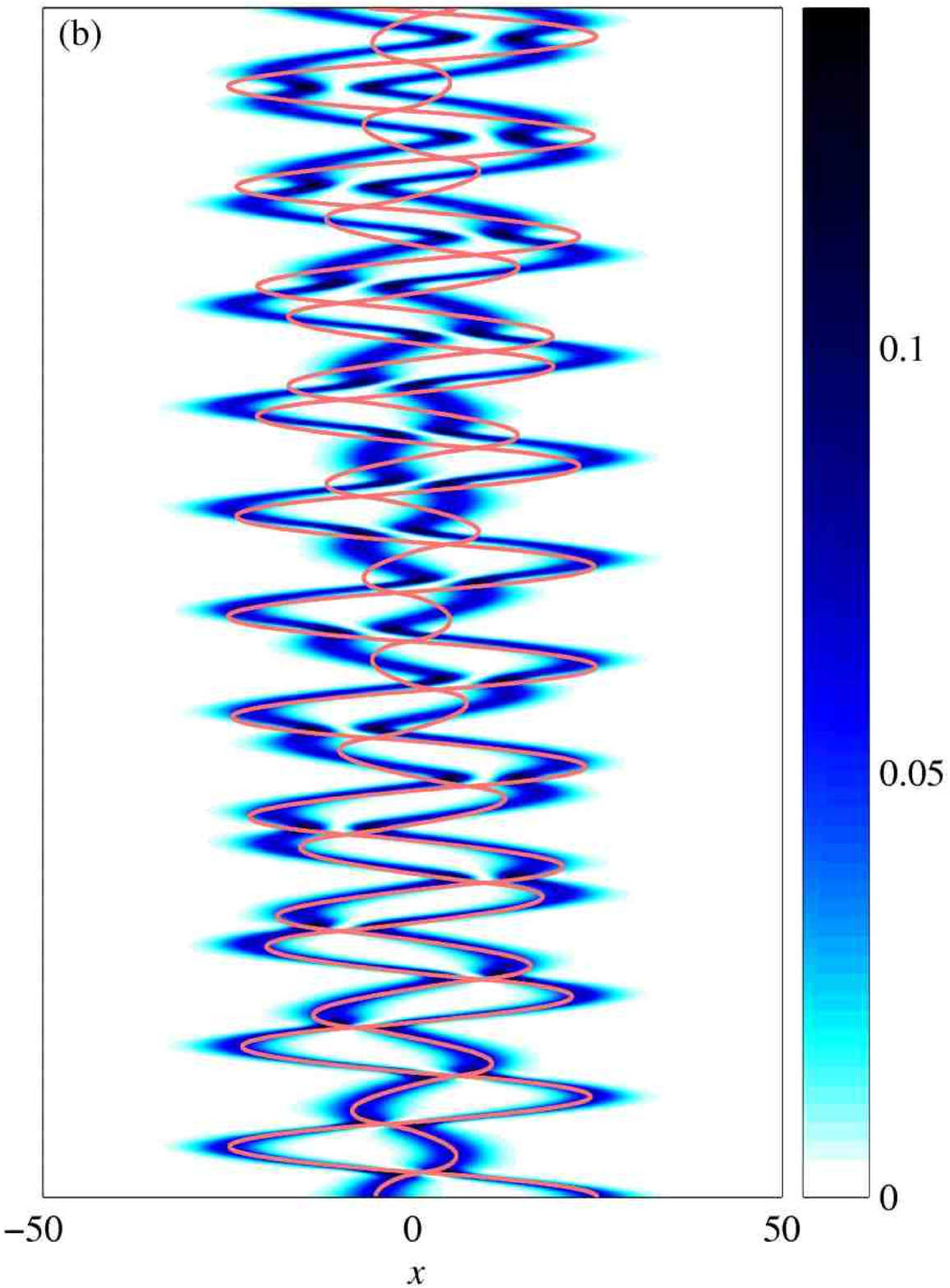}
\caption{(colour online). Trajectories in the particle model (lines) plotted over density distributions predicted by 1D GPE dynamics. The trajectories correspond to those given in figure \ref{fig_two_b}, but with additional centre-of-mass displacements. The trajectories also correspond to the upper trajectory marked on figure \ref{fig_poin_2}(c). The relative phase of the solitons in the wave dynamics is zero in figure (a), and $\pi$ in figure (b). The solitons have equal effective masses, the axial trapping frequency is $10/2\pi$ Hz, and the other parameters (radial trap frequency of $800/2\pi$ Hz, atomic species mass and scattering length of $^{7}$Li, and 5000 particles per soliton) are comparable to those in recent experiment \cite{Strecker_Nature_2002}. The unit of $x$ is then equal to 3.6 $\mu$m, and a unit of $t$ to 1.4 ms. } \label{fig_two_c}
\end{figure*}

\begin{figure*}[tbp]
\centering
\includegraphics[height=12cm]{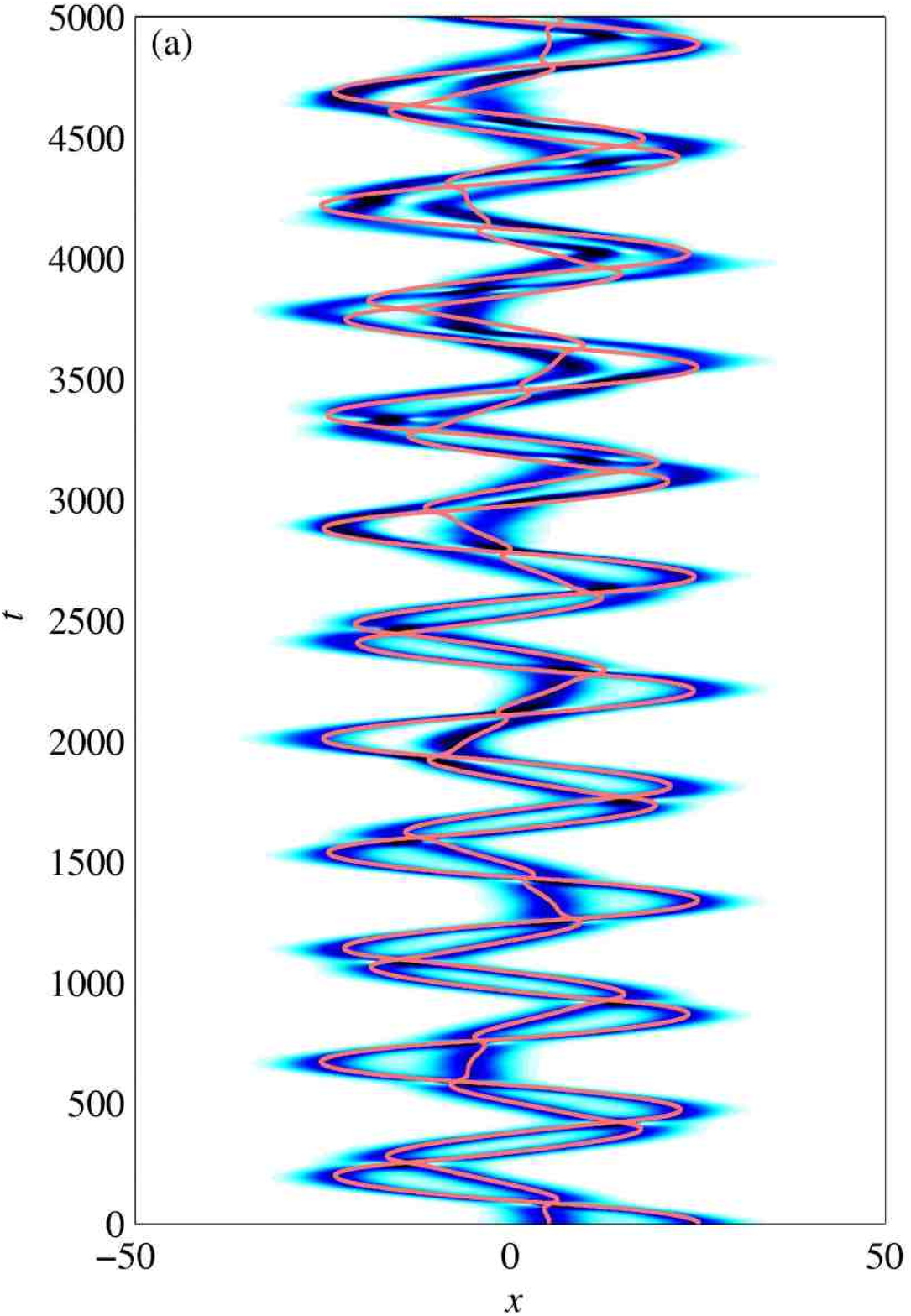}
\includegraphics[height=12cm]{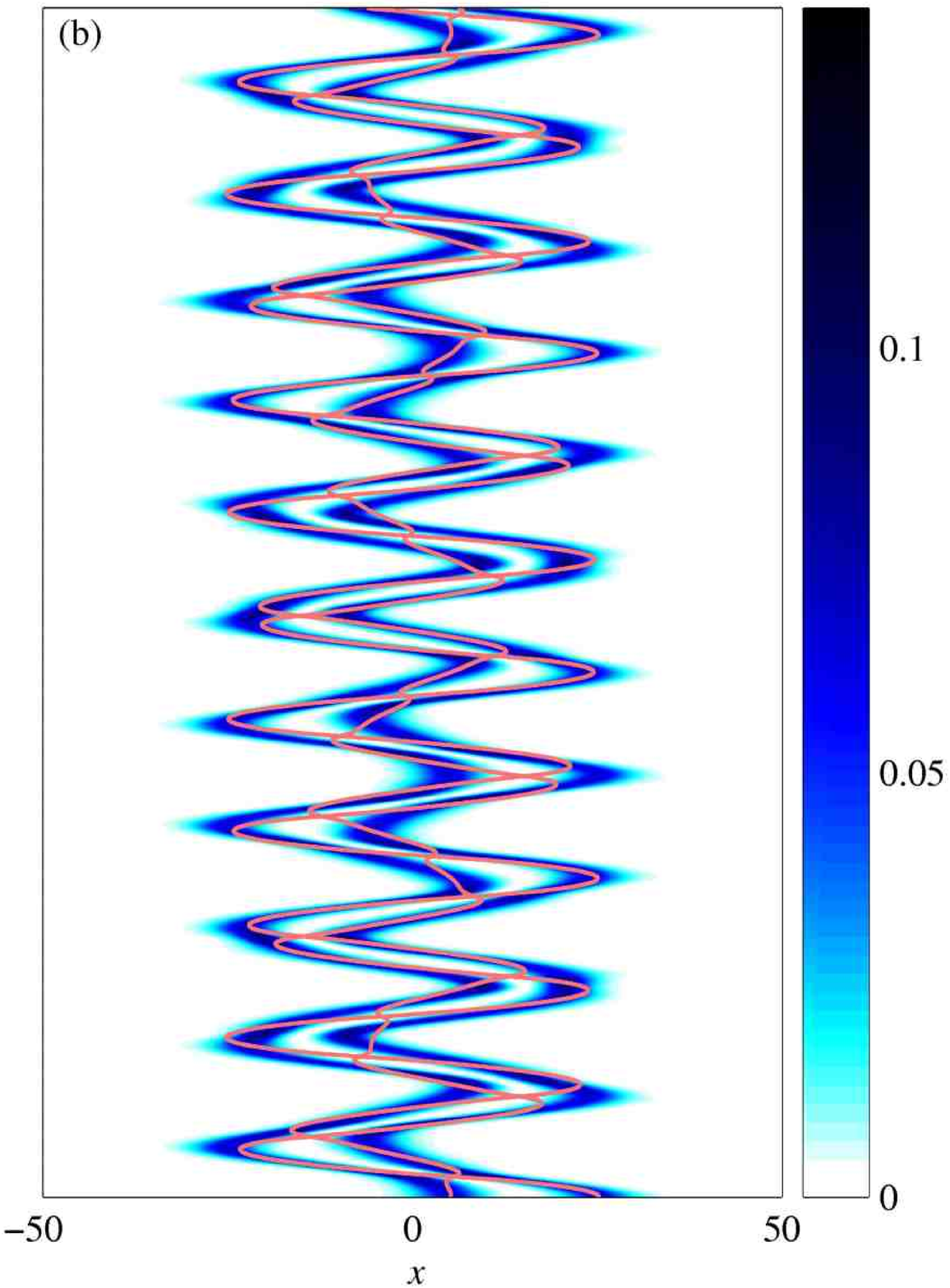}
\caption{(colour online). Trajectories in the particle model (lines) plotted over density distributions predicted by 1D GPE dynamics, corresponding to the trajectory marked on figure \ref{fig_poin_2}(d). The relative phase of the solitons in the wave dynamics is zero in figure (a), and $\pi$ in figure (b). The solitons have equal effective masses, the axial trapping frequency is  $10/2\pi$ Hz, and the other parameters (radial trap frequency of $800/2\pi$ Hz, atomic species mass and scattering length of $^{7}$Li, and 5000 particles per soliton) are comparable to those in recent experiment \cite{Strecker_Nature_2002}. The unit of $x$ is then equal to 3.6 $\mu$m, and a unit of $t$ to 1.4 ms. } \label{fig_two_c2}
\end{figure*}

\begin{figure*}[tbp]
\centering
\includegraphics[height=12cm]{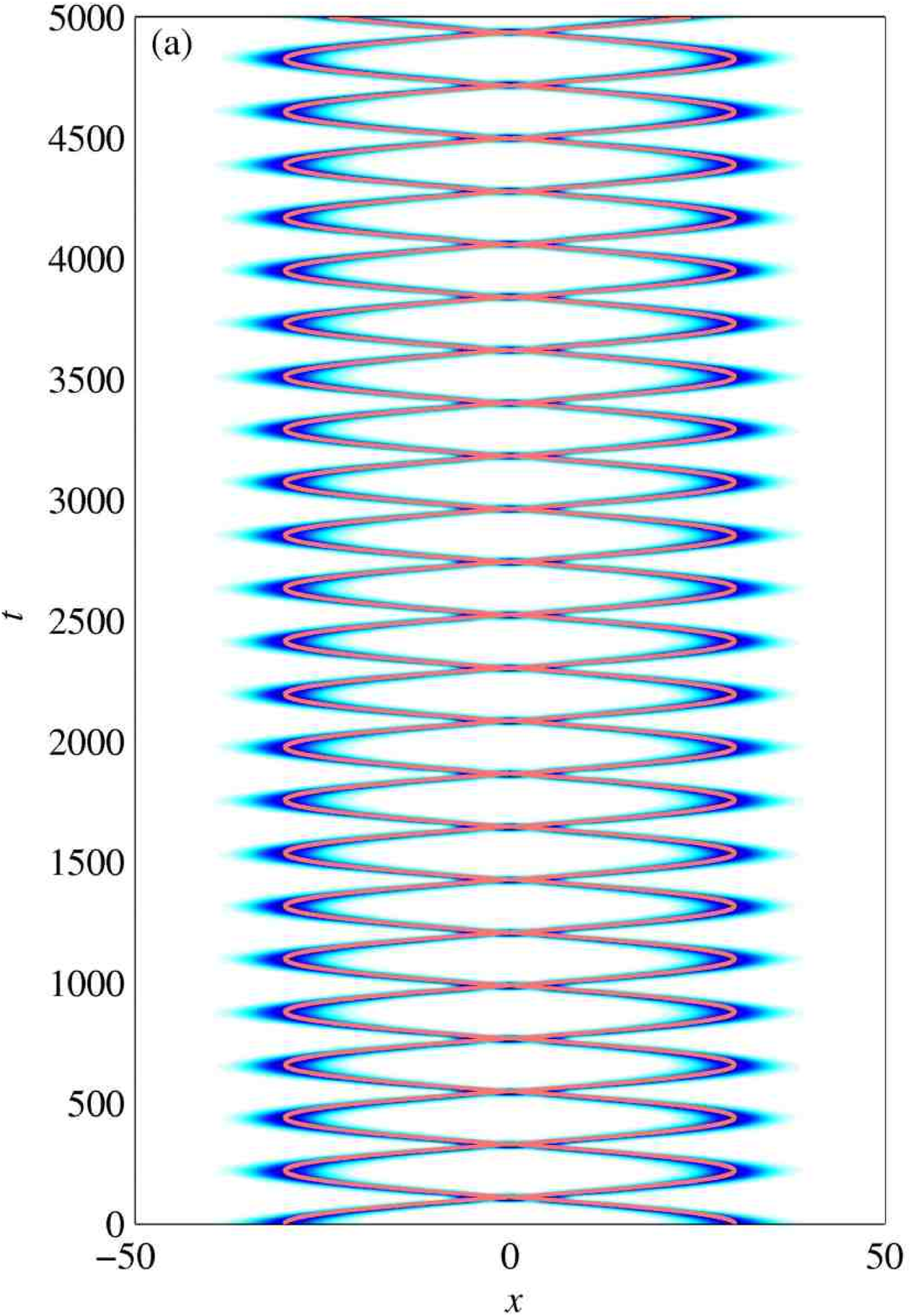}
\includegraphics[height=12cm]{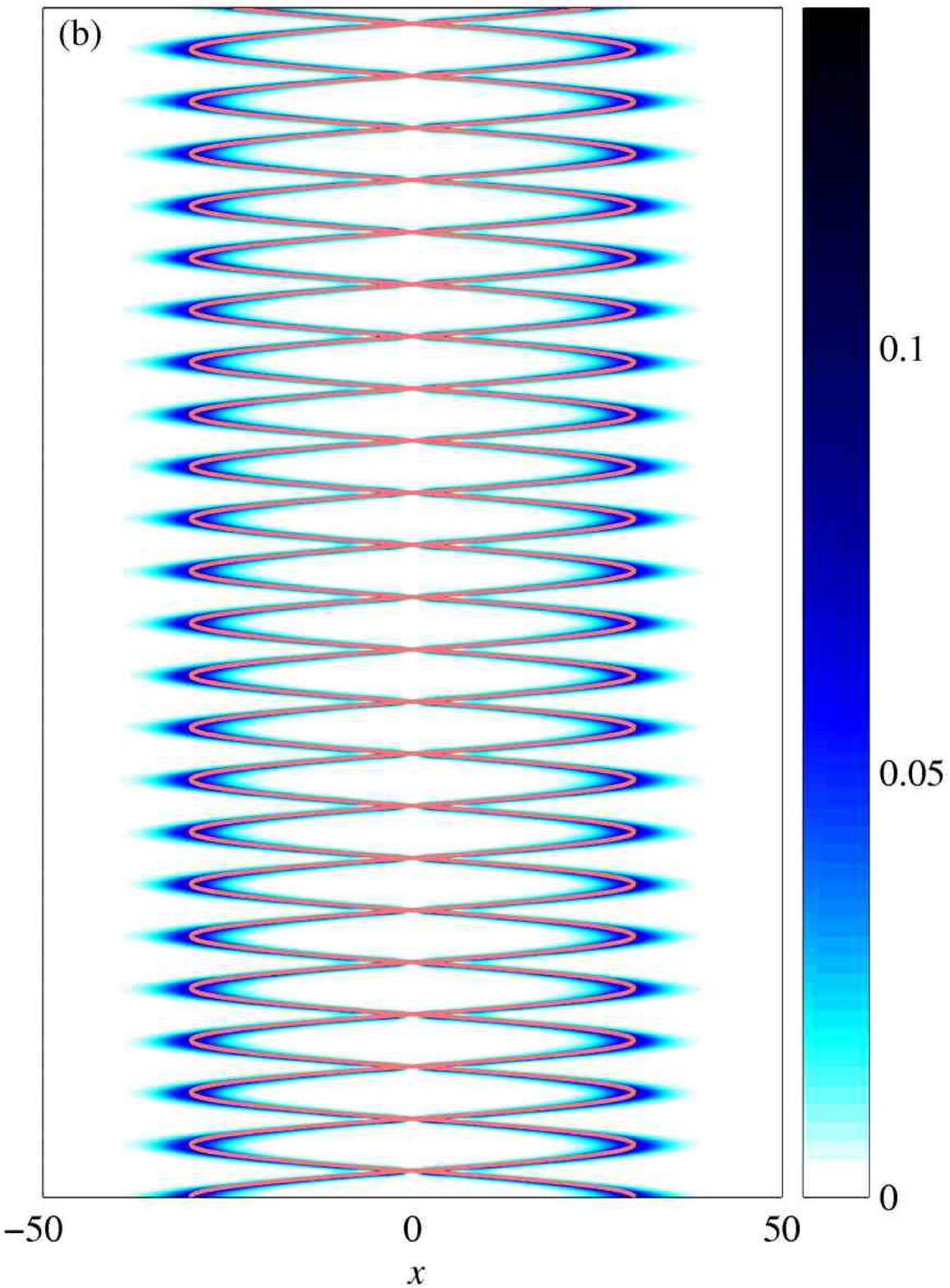}
\caption{(colour online). Trajectories in the particle model (lines) plotted over density distributions predicted by 1D GPE dynamics, corresponding to the trajectory marked on figure \ref{fig_poin_2}(d). The relative phase of the solitons in the wave dynamics is zero in figure (a), and $\pi$ in figure (b). The solitons have equal effective masses, the axial trapping frequency is  $10/2\pi$ Hz, and the other parameters (radial trap frequency of $800/2\pi$ Hz, atomic species mass and scattering length of $^{7}$Li, and 5000 particles per soliton) are comparable to those in recent experiment \cite{Strecker_Nature_2002}. The unit of $x$ is then equal to 3.6 $\mu$m, and a unit of $t$ to 1.4 ms.  } \label{fig_two_d}
\end{figure*}

\begin{figure}[tbp]
\centering
\includegraphics[width=6.0cm,angle=270]{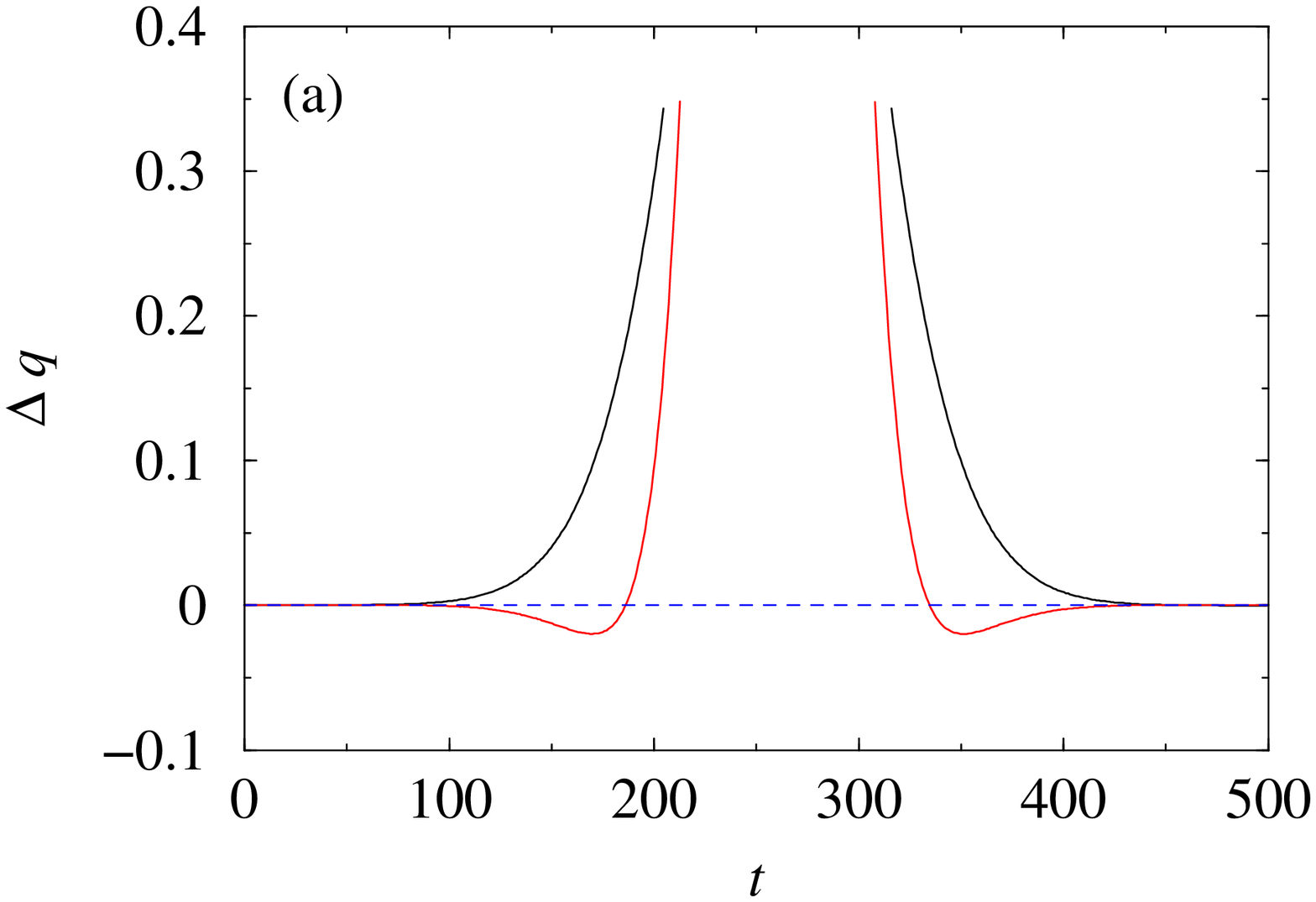}
\includegraphics[width=6.0cm,angle=270]{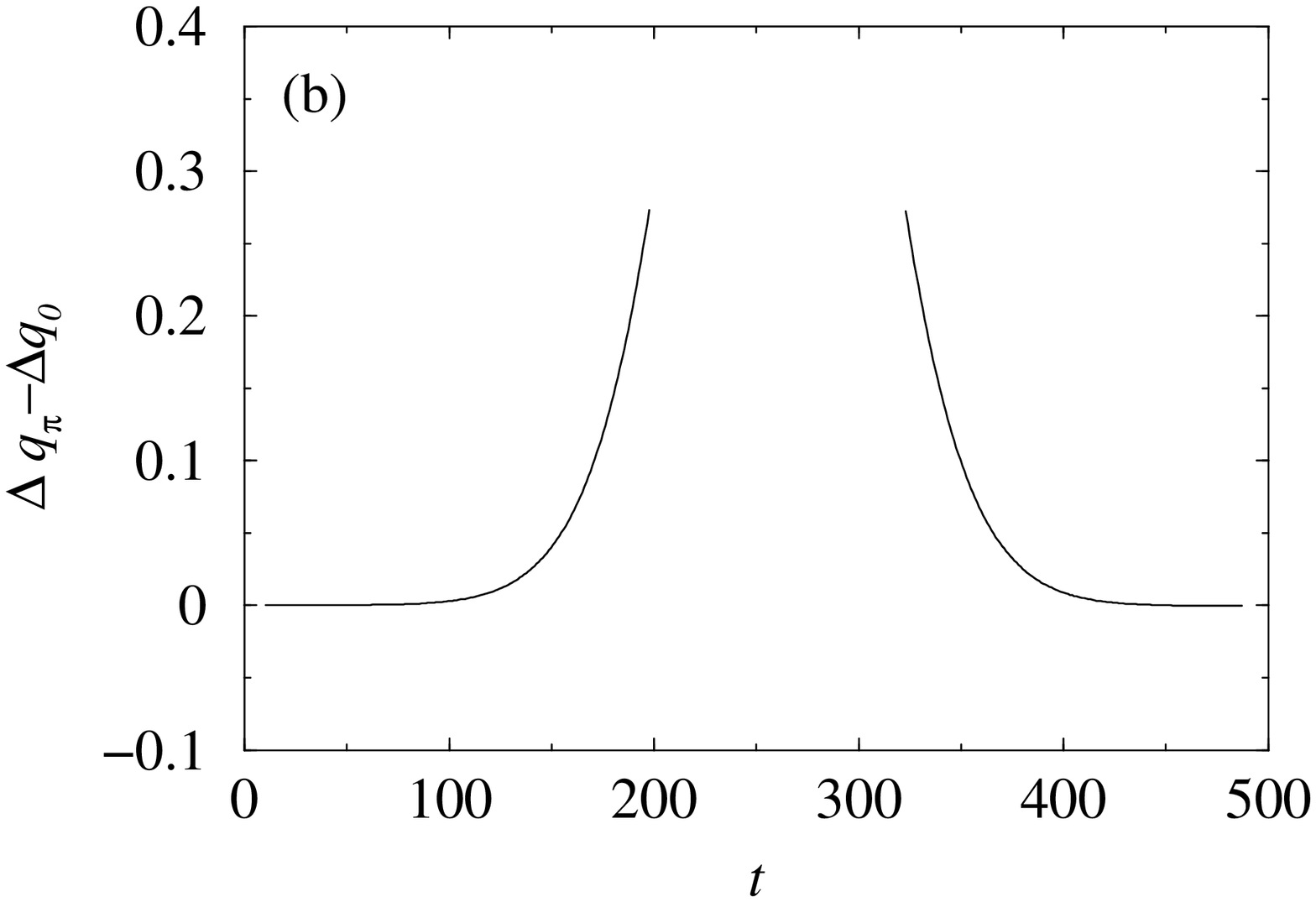}
\caption{(colour online) (a) Difference between soliton peak trajectory from wave simulation and particle model. The trajectories are those of the left-hand soliton in figure \ref{fig_phase1} going into and re-emerging from a collision in the case of $\Delta\phi=\pi$ (dark line) and $\Delta\phi=0$ (light line). Zero is indicated by the dotted line.  The difference $\Delta q$ is equal to the particle trajectory $q_{1}$ minus the position of the peak of the left-hand soliton, or equivalently the position of the peak of the right-hand soliton minus the particle trajectory $q_{2}$.
(b) Difference between the curves in figure (a). Note that the separation between the trajectories in the particle model and GPE dynamics is always larger in the $\pi$-phase case.
} \label{fig_DeltaP}
\end{figure}

In the Poincar\'e sections of Fig.\ \ref{fig_poin_2}, we highlighted a number of trajectories in white.  
These trajectories are plotted in position space as a function of time, overlaying density plots of corresponding 1D GPE solutions, in Figs.\ \ref{fig_two_b}--\ref{fig_two_d}.  We do this to test how accurately the particle model represents the GPE dynamics for a range of initial conditions.  

In particular, the wave-dynamics involve a phase variable not accounted for in the particle model. As shown in Fig.\ \ref{fig_phase1}, the phase difference between the solitons has an observable effect on the solitons' form during collisons, although as the solitons tend asymptotically apart, the solitons' density dynamics are insensitive to this.  Two solitons with equal norms in a harmonic trap have the same collisional form for all subsequent collisions; i.e., two solitons initially colliding with a phase difference $\phi_{\mbox{\scriptsize coll}}$, will have this phase difference for all subsequent collisions (see appendix \ref{App:CollisionalForm}).  This property allows results of GPE simulations with repeated in-phase and $\pi$ out-of-phase collisions to be compared for any trajectory in the two-particle model. Regimes with phase differences between zero and $\pi$ are not considered here, but will generally be expected to display behaviour intermediate between that of the zero and $\pi$ cases.

The trajectories displayed in Fig.\ \ref{fig_two_b} correspond to the fixed point (marked by a white star) in the Poincar\'e section of Fig.\ \ref{fig_poin_2}(b). We observe significantly better agreement between the particle model and the wave dynamics for the in-phase case [Fig.\ \ref{fig_two_b}(a)] than for the out-of-phase case [Fig.\ \ref{fig_two_b}(b)], where an obvious discrepancy between the particle and wave dynamics gradually accumulates.  Subsequent collisions consistently occur slightly earlier in the particle model [dynamics induced by the Hamiltonian (\ref{Ham_2S_identical})] than for the solitons propagated by the GPE [Eq.\ (\ref{S2})], so that by $t=5000$ quite a noticeable shift of the particle dynamics has taken place.  For the zero phase case, a similar systematic discrepancy is just observable at large $t$, in Fig.\ \ref{fig_two_b}(a); however, this discrepancy is very small and not readily apparent for most of the evolution. We note that $t=5000$ corresponds to 7 seconds using the parameters of Ref.\ \cite{Strecker_Nature_2002}, i.e., substantially longer than most typical experimental time scales. 

Figure \ref{fig_two_c} shows trajectories equivalent to those of Fig.\ \ref{fig_two_b}, but with an additional centre-of-mass motion. This corresponds to the upper of the two trajectories marked in white in the Poincar\'e section shown in Fig.\ \ref{fig_poin_2}(c), where there is gradual energy exchange between the solitons. We see the same discrepancy between particle and wave dynamics as was observed in Fig.\ \ref{fig_two_b}.  This is to be expected, because, as shown in appendix \ref{App:HarmonicallyOscillating}, incorporating a centre-of-mass oscillation into a solution of the harmonic GPE simply causes the wave-function's density profile to oscillate, without any further effect on its overall evolution.

The trajectories in Fig.\ \ref{fig_two_c2} correspond to the lower trajectory illustrated in the Poincar\'e section of Fig.\ \ref{fig_poin_2}(c), where there is rapid energy exchange between the solitons. In this regime, good agreement is found in the case of in-phase collisions. For the case of out-of-phase collisions, there is some degree of agreement, but there are obvious discrepancies.  Notably, the solitons in the GPE appear to remain a fixed distance apart, whereas the particle trajectories necessarily show repeated collisions.

The dynamics shown in Fig.\ \ref{fig_two_d} correspond to the fixed point (marked by a white star) in the Poincar\'e section of Fig.\ \ref{fig_poin_2}(d). In this regime, the solitons collide with a significantly faster relative speed than in Figs. \ref{fig_two_b}--\ref{fig_two_c2}, and the collision time is accordingly shorter compared to the trap period. In this regime it is clear that the agreement between the particle model and the wave dynamics is much better than in the previous cases (figures \ref{fig_two_b} to \ref{fig_two_c2}), with the relative phase having much less of an observable influence.

\subsubsection{Discussion of results comparing GPE to particle evolutions  \label{Sec:Discussion}}
To explain the observed discrepancies between corresponding particle and wave evolutions, particularly when the solitons collide out-of-phase (see Figs. \ref{fig_two_b} and  \ref{fig_two_c}), we must consider the assumptions made while composing the effective particle Hamiltonian [Eq.\ (\ref{Ham4})]. When constructing this Hamiltonian, we stated that the collision time should be small compared with the trap period. To characterise the timescale of a collision, it is instructive to consider how rapidly the trajectories in the particle model converge, to those deduced from motion of the soliton peaks in the GPE, after a collision. 

We consider again the evolutions depicted in Fig.\ \ref{fig_phase1}(a) and \ref{fig_phase1}(b).  Figure \ref{fig_DeltaP} shows the differences in position between the peak of the left-hand soliton evolved by the GPE, and the corresponding trajectory in the particle model during a collision. As there is no external potential, the particle trajectories are asymptotically exact as $t\rightarrow \pm\infty$. It is clear that, in the case of an in-phase collision, the convergence of the particle and wave trajectories is more rapid than for a $\pi$ out-of-phase collision. 

In a harmonic trap, subsequent to a collision, two solitons can only move a finite distance apart (i.e., \textit{not\/} asymptotically far) before moving together and colliding once more.  The dynamics of solitons colliding with a $\pi$ phase difference are therefore not expected to agree as well with the effective particle model dynamics, compared with solitons colliding in-phase.  This effect builds up over time because the phase difference is preserved for repeated collisions.   Figure \ref{fig_DeltaP} shows that, in the $\pi$ out-of-phase case, compared to the particle trajectory, the trajectory of the soliton peak tends to be further away from the point of collision, with the two trajectories converging asymptotically as $t$ tends to a time infinitely before and after the time of collision.  Due to the harmonic potential, the particle trajectories consequently start their return to the centre of the trap at points closer to the centre, compared to the $\pi$ out-of-phase solitons.  This explains why subsequent collisions take place earlier for the particle trajectories than predicted by GPE dynamics, as observed in Fig.\ \ref{fig_two_b}.  For large approach speeds (Fig.\ \ref{fig_two_d}), however, the collision time is sufficiently small for both in-phase and out-of-phase cases, such that any discrepancy between the predictions of the GPE and the effective particle model is too small to be observed.

Figure \ref{fig_two_c2} represents a regime where the solitons do not separate well between collisions. The particle model might not be expected to apply well to such regimes; nevertheless, reasonably good agreement is observed in Fig.\ \ref{fig_two_c2}(a), with somewhat less good agreement (as expected) for the $\pi$ out-of-phase case shown in Fig.\ \ref{fig_two_c2}(b). The density distribution of figure \ref{fig_two_c2}(b) is essentially a continuous collision of two solitons with a $\pi$ phase-difference, and has the appearance of two individual wavepackets that never cross.  These are expected to be better described by alternative treatments, described by Gordon \cite{Gordon_OptLett_1983} in the absence of any external potential, and by Gerdjikov \textit{et al.} \cite{Gerdjikov_PRE_2006} in the case of harmonic confinement, or may alternatively be treated by perturbative approaches  \cite{Karpman_Physica3D_1981,Kivshar_RMP_1989,Okamawari_PRA_1995}. 
 
In 3D simulations, collisions of solitons with a $\pi$ phase difference have been predicted to have some degree of immunity against collapse, as opposed to in-phase collisions \cite{Parker_unpublished_2006,Carr_PRL_2004}. It is clear from the above analysis that systems of two solitons of equal norm will be able to maintain a $\pi$ phase difference for all repeated collisions for any choice of initial positions and momenta, provided the right initial phase difference is chosen, and thus remain stable.  Systems of two solitons with different norms have not been studied here, but are an area for future study; we recall that the particle model is expected to remain valid for regimes of different norms as long as $ 2|\eta_{1}-\eta_{2}|\ll |v_{1}-v_{2}|$, as explained in Sec.\ \ref{Sec:ParticleModel}. 

\subsection{Three solitons \label{ThreeSolitons}}

\subsubsection{Particle model \label{ThreeParticleModel}} 

\begin{figure*}[tbp]
\centering
\includegraphics[width=6.0cm,angle=270]{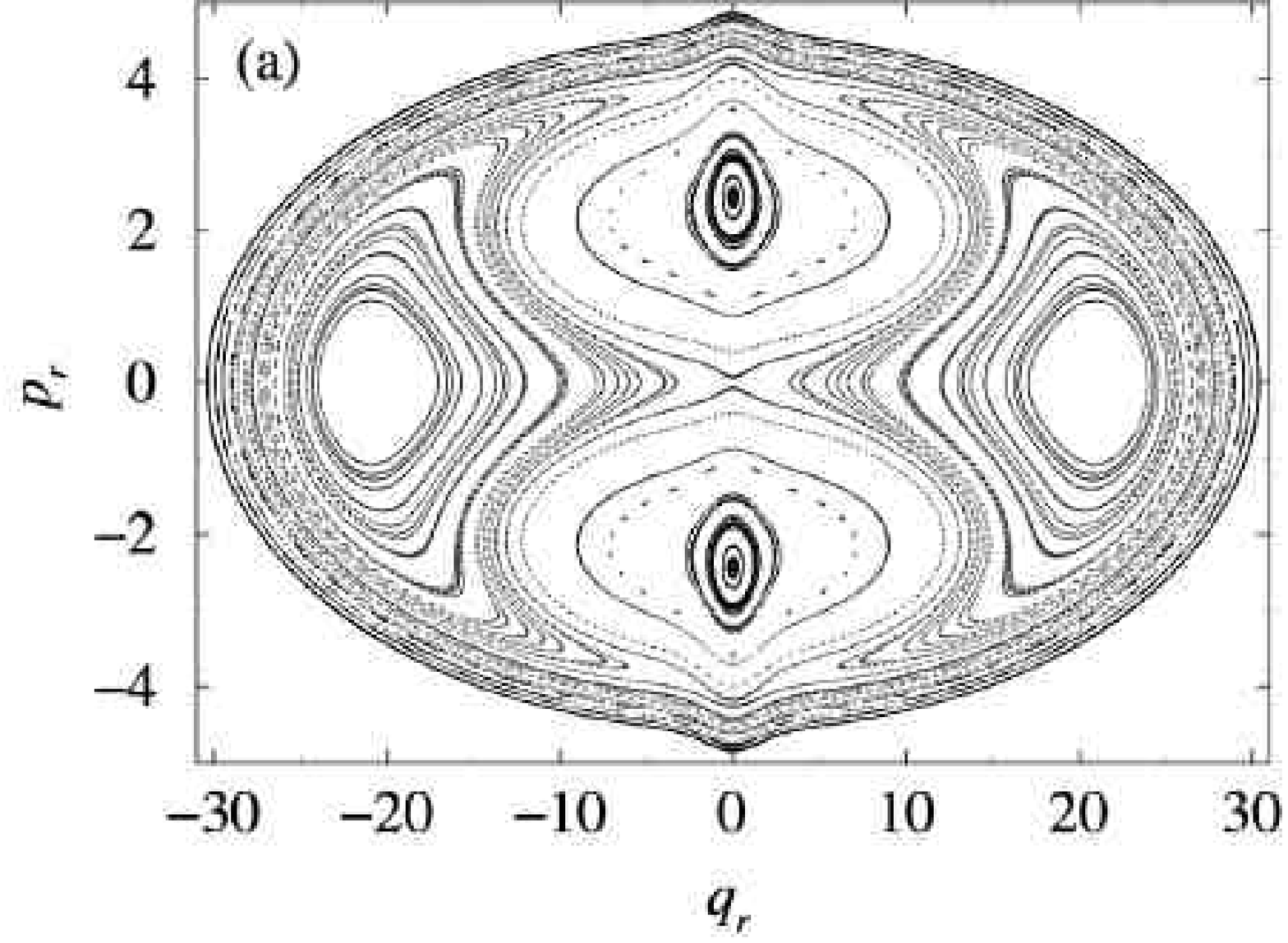}
\quad
\includegraphics[width=6.0cm,angle=270]{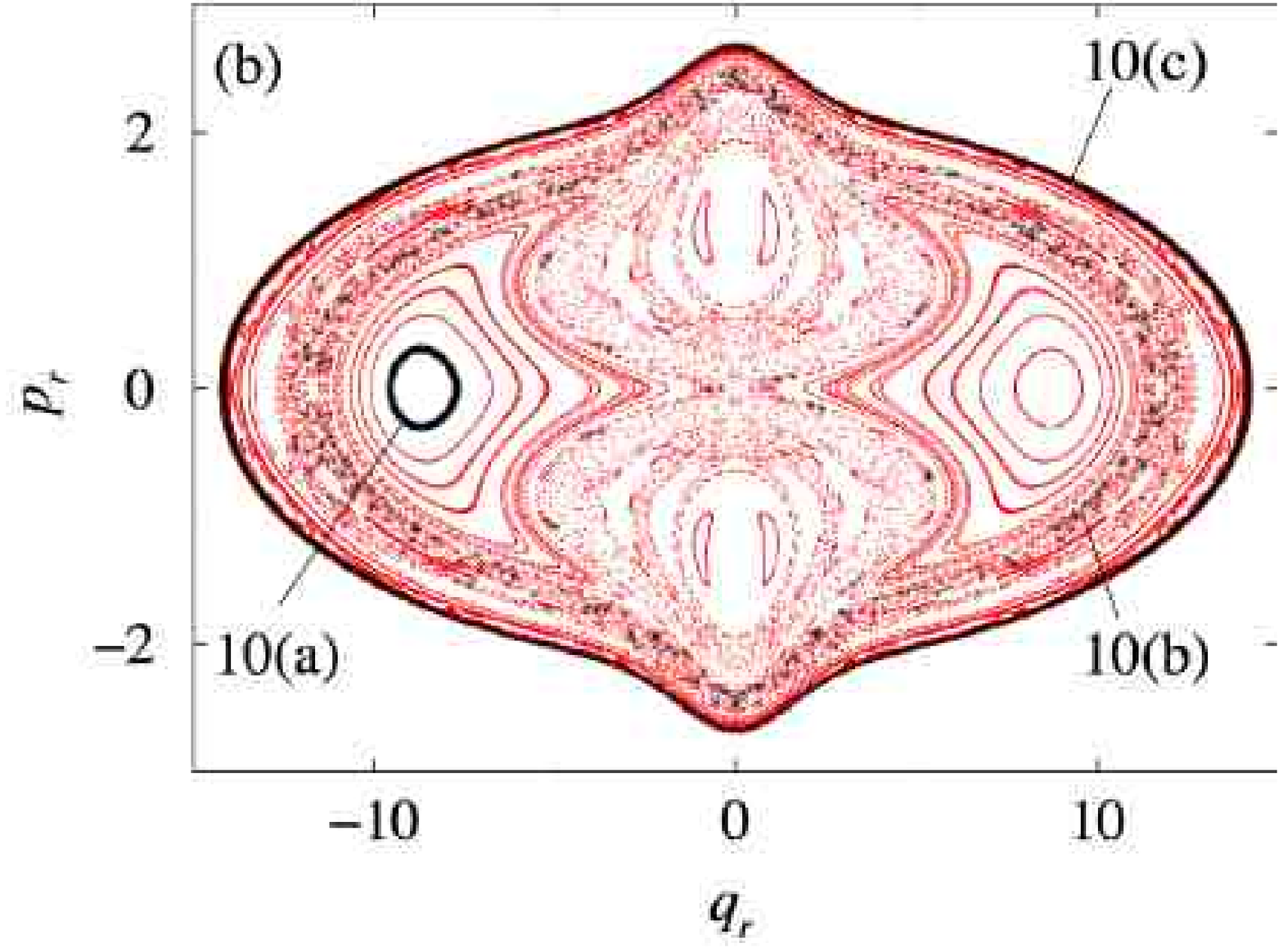}
\includegraphics[width=6.0cm,angle=270]{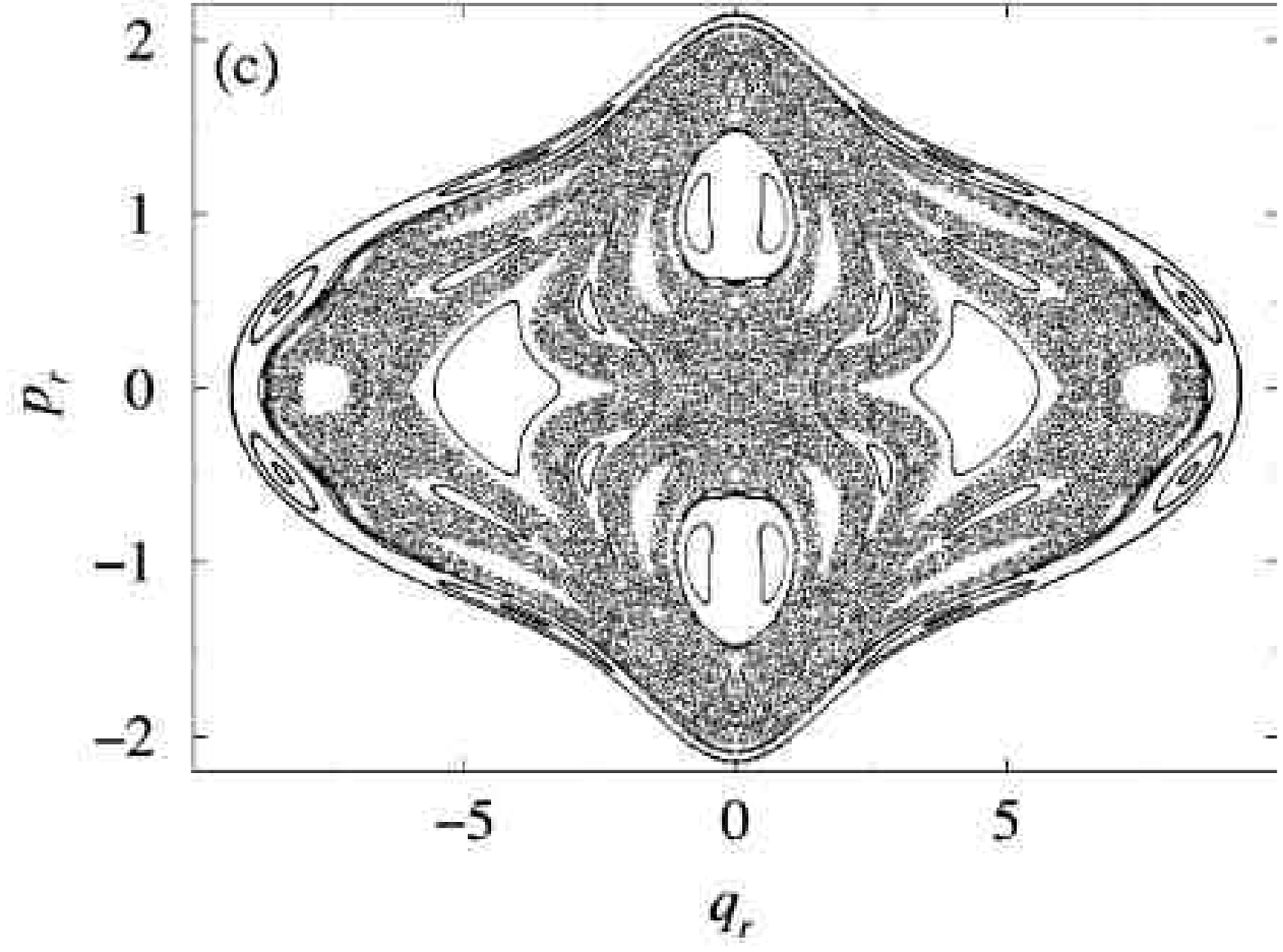}
\quad
\includegraphics[width=6.0cm,angle=270]{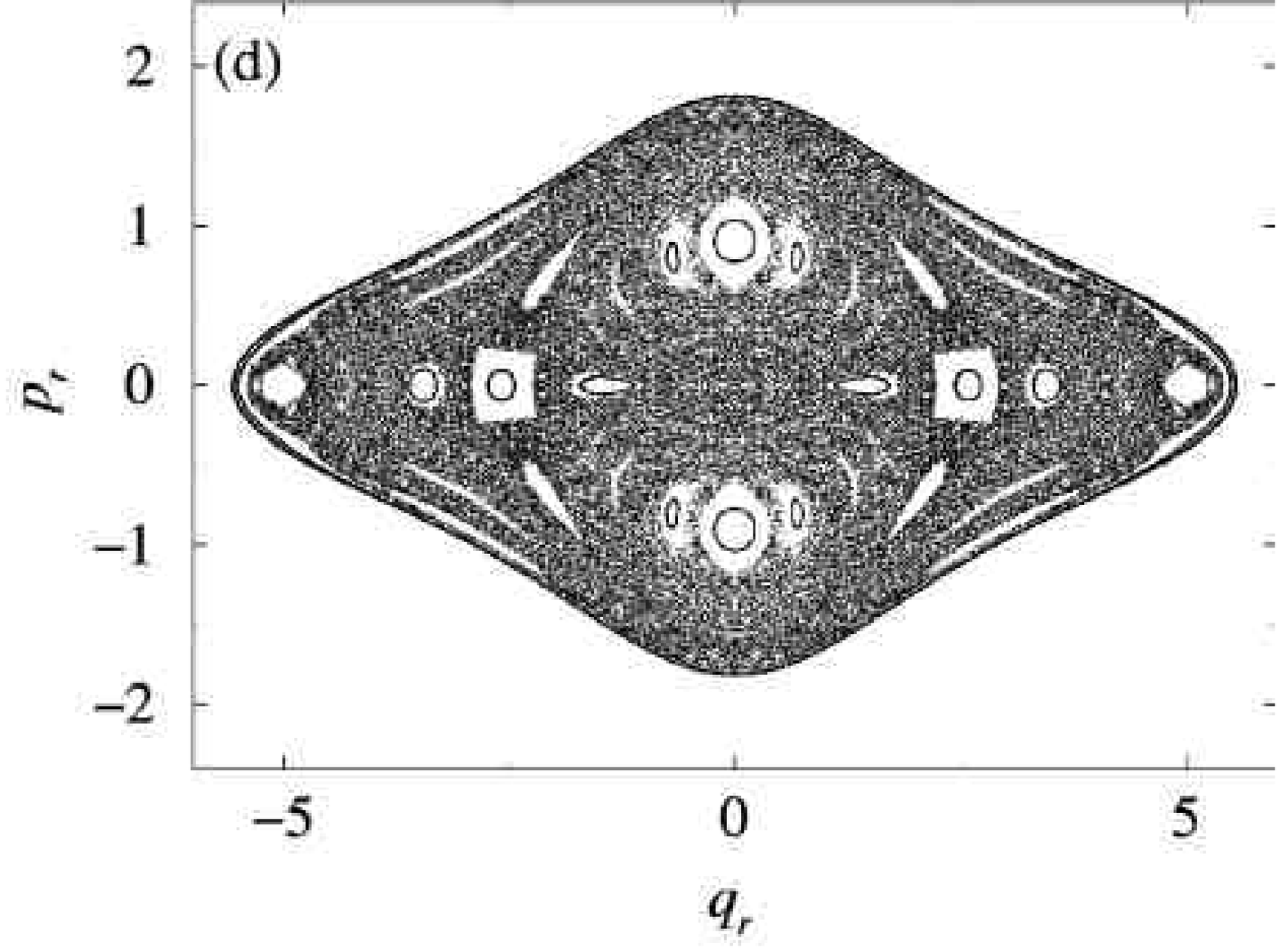}
\includegraphics[width=6.0cm,angle=270]{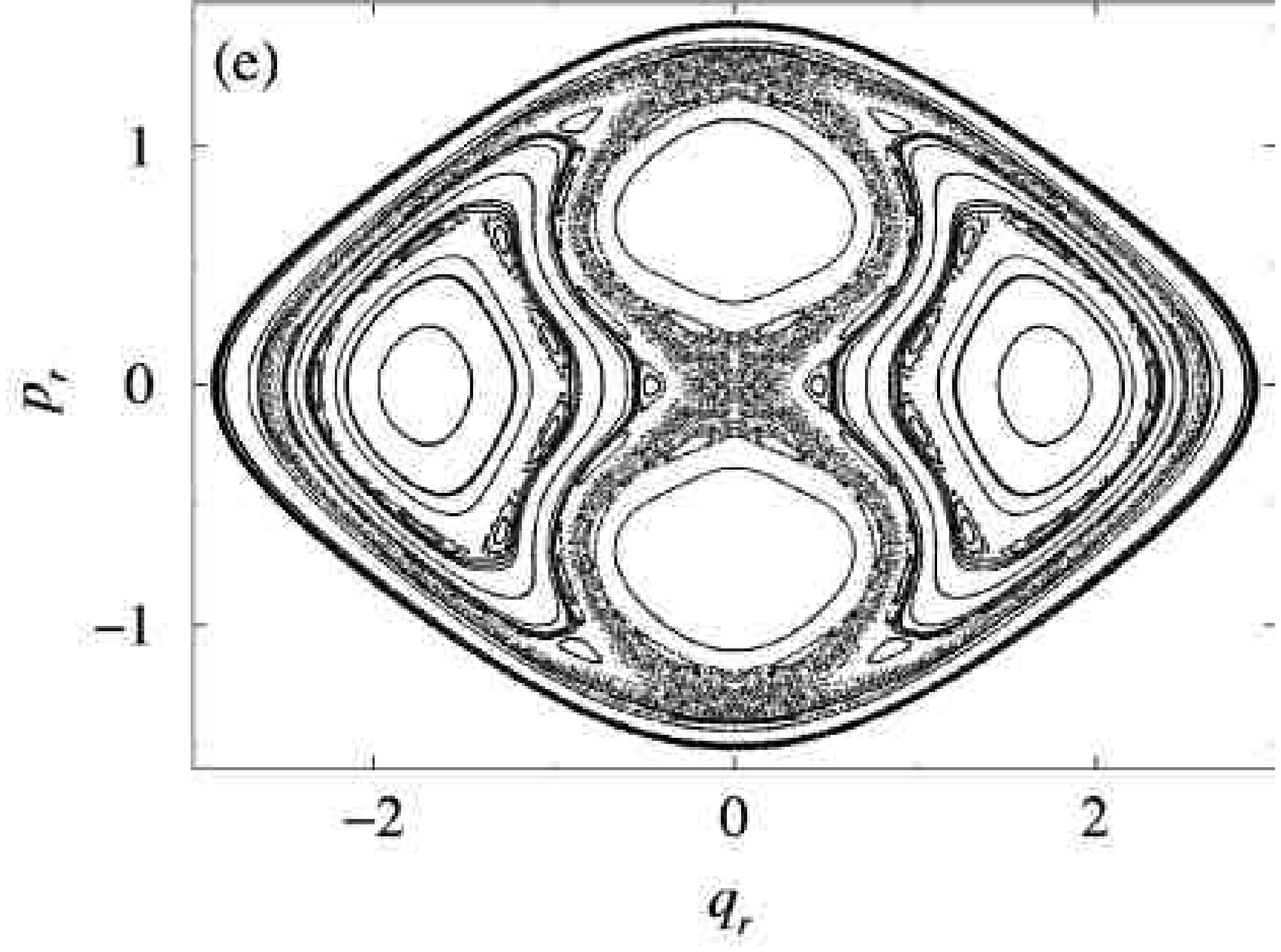}
\quad
\includegraphics[width=6.0cm,angle=270]{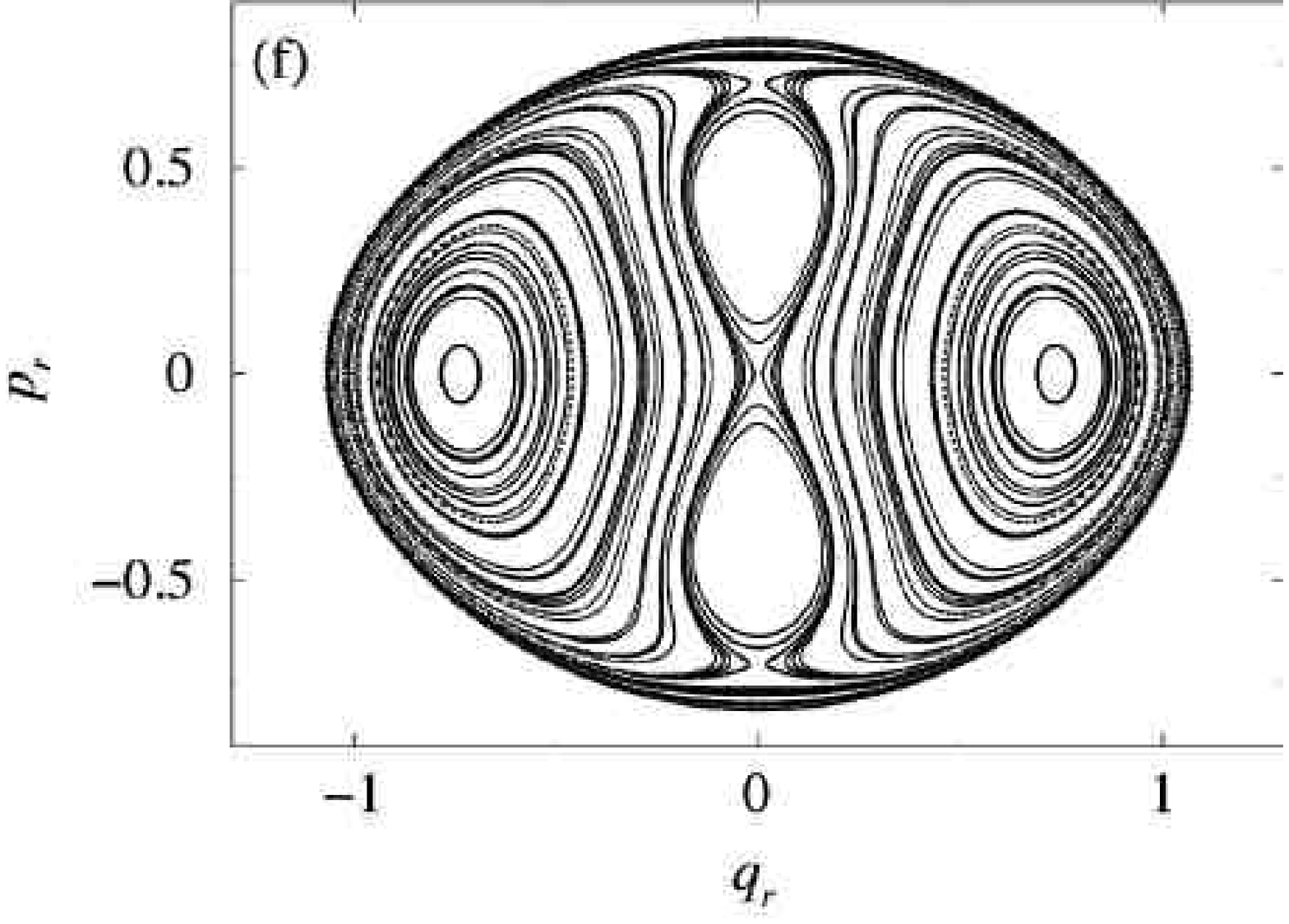}
\caption{(Colour online).  Poincar\'e section of the three-soliton system with (a) $\tilde{H}=60$; (b) $\tilde{H}=10$, regions corresponding to trajectories in figures 2(a) to 2(c) are labeled, and highlighted using larger, darker points; (c) Poincar\'e section of the system with $\tilde{H}=2$; (d) $\tilde{H}=-2$; (e) $\tilde{H}=-5$; (f) $\tilde{H}=-10$
The section corresponds to the momentum $p_{r}$ and position $q_{r}$ of the ``asymmetric stretch''
mode when the ``stretch'' mode coordinates $q_{c}=0$, $p_{c}<0$.
The figures correspond to the regime where the solitons have equal effective masses, the axial trapping frequency is $10/2\pi$ Hz, and the other parameters (radial trap frequency of $800/2\pi$ Hz, atomic species mass and scattering length of $^{7}$Li, and 5000 particles per soliton) correspond to recent experiment \cite{Strecker_Nature_2002}.} \label{fig_poin}
\end{figure*}

Whereas for two solitons, the particle model dynamics are always regular (see Sec.\ \ref{Sec:ParticleModel}), in the case of three solitons ($N_{s}=3$), the situation is quite different. A useful coordinate system for the three soliton system is to be found in the normal coordinates of the system for small displacements of the particles from the origin: the centre-of-mass position
\begin{equation}
Z_{T}:=\frac{\eta_{1}q_{1}+\eta_{2}q_{2}+\eta_{3}q_{3}}{\eta_{1}+\eta_{2}+\eta_{3}},
\end{equation}
\begin{equation} 
z_{c}:=\frac{\eta_{1}(\eta_{2}+2\eta_{3})q_{1}+\eta_{2}(\eta_{3}-\eta_{1})q_{2}-\eta_{3}(\eta_{2}+2\eta_{1})q_{3}}{\eta_{1}\eta_{2}+\eta_{2}\eta_{3}+4\eta_{1}\eta_{3}},
\end{equation} (corresponding to the ``stretch'' mode), and 
\begin{equation}
z_{r}:=q_{1}-2q_{2}+q_{3},
\end{equation} (corresponding to the ``asymmetric stretch'').  The stretch modes are similar to those used to describe vibrational dynamics in a tri-atomic molecule \cite{Tennyson_ChemPhys_1985}; as the system is constrained to 1D, however, there is no analogue of the molecular bending mode.  Using these coordinates, the three-particle Hamiltonian [Eq.\ (\ref{Ham4})] takes the form:
\begin{widetext}
\begin{equation}
\begin{split}
H= & \frac{1}{2}\left[ \frac{W_{T}^{2}}{\eta_{1}+\eta_{2}+\eta_{3}}+w_{c}^{2}\frac{\eta_{1}+\eta_{2}+\eta_{3}}{\eta_{1}\eta_{2}+\eta_{2}\eta_{3}+4\eta_{1}\eta_{3}} +w_{r}^{2}\frac{\eta_{1}\eta_{2}+\eta_{2}\eta_{3}+4\eta_{1}\eta_{3}}{\eta_{1}\eta_{2}\eta_{3}}\right]
\\&+\frac{\omega^{2}}{2}\left[Z_{T}^{2}(\eta_{1}+\eta_{2}+\eta_{3})+z_{c}^{2}\frac{\eta_{1}\eta_{2}+\eta_{2}\eta_{3}+4\eta_{1}\eta_{3}}{\eta_{1}+\eta_{2}+\eta_{3}}+z_{r}^{2}\frac{\eta_{1}\eta_{2}\eta_{3}}{\eta_{1}\eta_{2}+\eta_{2}\eta_{3}+4\eta_{1}\eta_{3}}\right]
\\ &
-2\eta_{1}\eta_{2}(\eta_{1}+\eta_{2})\mathrm{sech}^{2}\left[\frac{2\eta_{1}\eta_{2}}{\eta_{1}+\eta_{2}}\left(\frac{\eta_{2}\eta_{3}+2\eta_{1}\eta_{3}}{\eta_{1}\eta_{2}+\eta_{2}\eta_{3}+4\eta_{1}\eta_{2}} \right)z_{r}+z_{c}\right]
\\&-2\eta_{1}\eta_{3}(\eta_{1}+\eta_{3})\mathrm{sech}^{2}\left[\frac{2\eta_{1}\eta_{3}}{\eta_{1}+\eta_{3}}\left(\frac{\eta_{2}\eta_{3}-2\eta_{1}\eta_{2}}{\eta_{1}\eta_{2}+\eta_{2}\eta_{3}+4\eta_{1}\eta_{2}} \right)z_{r}+2z_{c}\right]
\\&-2\eta_{2}\eta_{3}(\eta_{2}+\eta_{3})\mathrm{sech}^{2}\left[\frac{2\eta_{2}\eta_{3}}{\eta_{2}+\eta_{3}}\left(\frac{-(\eta_{1}\eta_{2}+2\eta_{1}\eta_{3})}{\eta_{1}\eta_{2}+\eta_{2}\eta_{3}+4\eta_{1}\eta_{2}} \right)z_{r}+z_{c}\right], \label{Ham_three}
\end{split}
\end{equation}
\end{widetext}
where $W_{T}=p_{1}+p_{2}+p_{3}$, $w_{c} = [(\eta_{2}+2\eta_{3})p_{1}+(\eta_{3}-\eta_{1})p_{2}-(\eta_{2}+2\eta_{1})p_{3}]/(\eta_{1}+\eta_{2}+\eta_{3})$, and $w_{r}= (\eta_{2}\eta_{3}p_{1} - 2 \eta_{1}\eta_{3} p_{2} + \eta_{1}\eta_{2} p_{3})/(\eta_{1}\eta_{2} + \eta_{2}\eta_{3} + 4\eta_{1}\eta_{3})$ are the momenta canonically conjugate to the coordinates $Z_{T}$, $z_{c}$, and $z_{r}$, respectively.  It is apparent in Eq.\ (\ref{Ham_three}) that the Hamiltonian, as in the two-particle case, is decoupled into a centre-of-mass component, and a component describing the stretch modes (which are coupled to each other).

In the case of identical effective masses, the coordinates simplify substantially.  It turns out to be convenient to consider slightly different coordinates, however, as this produces a simpler final form for the Hamiltonian describing the stretch mode dynamics.  We therefore define $Q_{T}= \eta Z_{T}= \eta(q_{1}+q_{2}+q_{3})/3$, $q_{c}=\eta z_{c} = \eta(q_{1}-q_{3})/2$ and
$q_{r}=\eta z_{r} = \eta(q_{1}+q_{3}-2q_{2})$. We rescale the time to
$\tilde{t}=\eta^{2}t$, and then introduce the momenta $p_{c}=w_{c}/\eta^{2}=(p_{1}-p_{3})/\eta^{2}$ and $p_{r}=w_{r}/\eta^{2} = (p_{1}-2p_{2}+p_{3})/6\eta^{2}$.  Using these dynamical variables, the resultant Hamiltonian (the reduced system Hamiltonian), with the centre-of-mass degrees of freedom removed, becomes:
\begin{equation}
\begin{split}
\tilde{H}= & 3p_{r}^{2}+\frac{\omega^{2}}{2\eta^{4}}\frac{q_{r}^{2}}{12}+\frac{p_{c}^{2}}{4}+\frac{\omega^{2}}{2\eta^{4}}q_{c}^{2}-4\mathrm{sech}^{2}(2q_{c})
\\ &
-4\mathrm{sech}^{2}(q_{c}+\frac{q_{r}}{2})-4\mathrm{sech}^{2}(q_{c}-\frac{q_{r}}{2}). \label{Ham_red}
\end{split}
\end{equation}
This Hamiltonian, describing the two remaining degrees of freedom, is not separable, and it is necessary to integrate the corresponding Hamilton's equations of motion numerically to analyse the system's behaviour. As they represent a slice through the phase space of a system, Poincar\'e sections provide a good illustration of regions of regular and chaotic dynamics. In regions of regular behaviour, any trajectory will lie on a torus in phase-space, and will thus trace a closed curve in the Poincar\'e section; in regions of chaotic behaviour, a trajectory will go through every point in that region of phase space, and thus fill an area on the Poincar\'e section (a so-called ergodic sea) \cite{Gutzwiller_Book_1990,Reichl_Book_1992}.   We choose to show sections corresponding to the momentum $p_{r}$ and position $q_{r}$ of the ``asymmetric stretch'' mode when the ``stretch'' mode coordinate takes the value $q_{c}=0$, and when its canonically conjugate momentum $p_{c}<0$.  Other sections can be expected to be equally illustrative of the qualitative behaviour.

Figure \ref{fig_poin} shows six Poincar\'e sections for three different reduced system energies $\tilde{H}$. The behaviour is regular at large positive values of $\tilde{H}$ [Fig.
\ref{fig_poin} (a)], but as $\tilde{H}$ is reduced, chaotic behaviour emerges, characterised by ergodic regions in between regular tori. For small (negative) $\tilde{H}$ the system is mostly an ergodic sea, with islands of stability [Fig. \ref{fig_poin} (c)]; but as $\tilde{H}$ is made more negative, the chaotic regions begin to subside, and the behaviour becomes increasingly regular again. 

Consideration of the form of the reduced-system
Hamiltonian [Eq. (\ref{Ham_red})] shows that without the interaction the system is integrable, as it becomes a decoupled pair of harmonic oscillators. When $\tilde{H}$ is large and positive, the interaction part of the Hamiltonian (which is always negative) should give a relatively small contribution to the Hamiltonian, compared to the integrable part of the Hamiltonian (which is always positive). When $\tilde{H}$ is reduced, this is no longer the case, and chaotic dynamics are manifest. However, in regimes where the coordinates and momenta are close to zero, i.e., $\tilde{H}$ approaches its lower bound of $-12$, the interaction potential becomes approximately harmonic. The Hamiltonian $\tilde{H}$ takes the following seperable form:
\begin{equation}
\begin{split}
\tilde{H}= & 3p_{r}^{2}+\left(\frac{\omega^{2}}{24\eta^{4}}+2\right)q_{r}^{2}+\frac{p_{c}^{2}}{4}+\left(\frac{\omega^{2}}{2\eta^{4}}+24\right)q_{c}^{2},
\label{Ham_red_2}
\end{split}
\end{equation}
i.e., it again describes a pair of decoupled harmonic oscillators.  We consequently expect the phase-space structure to be qualitatively similar in the opposing limits of $\tilde{H}$ very large and positive, and $\tilde{H}$ large and negative. From Figs.\ \ref{fig_poin}(a) and \ref{fig_poin}(f), we do indeed observe this to be the case.

\subsubsection{Comparison with GPE simulations}

\begin{figure*}[tbp]
\centering
\includegraphics[height=11.5cm]{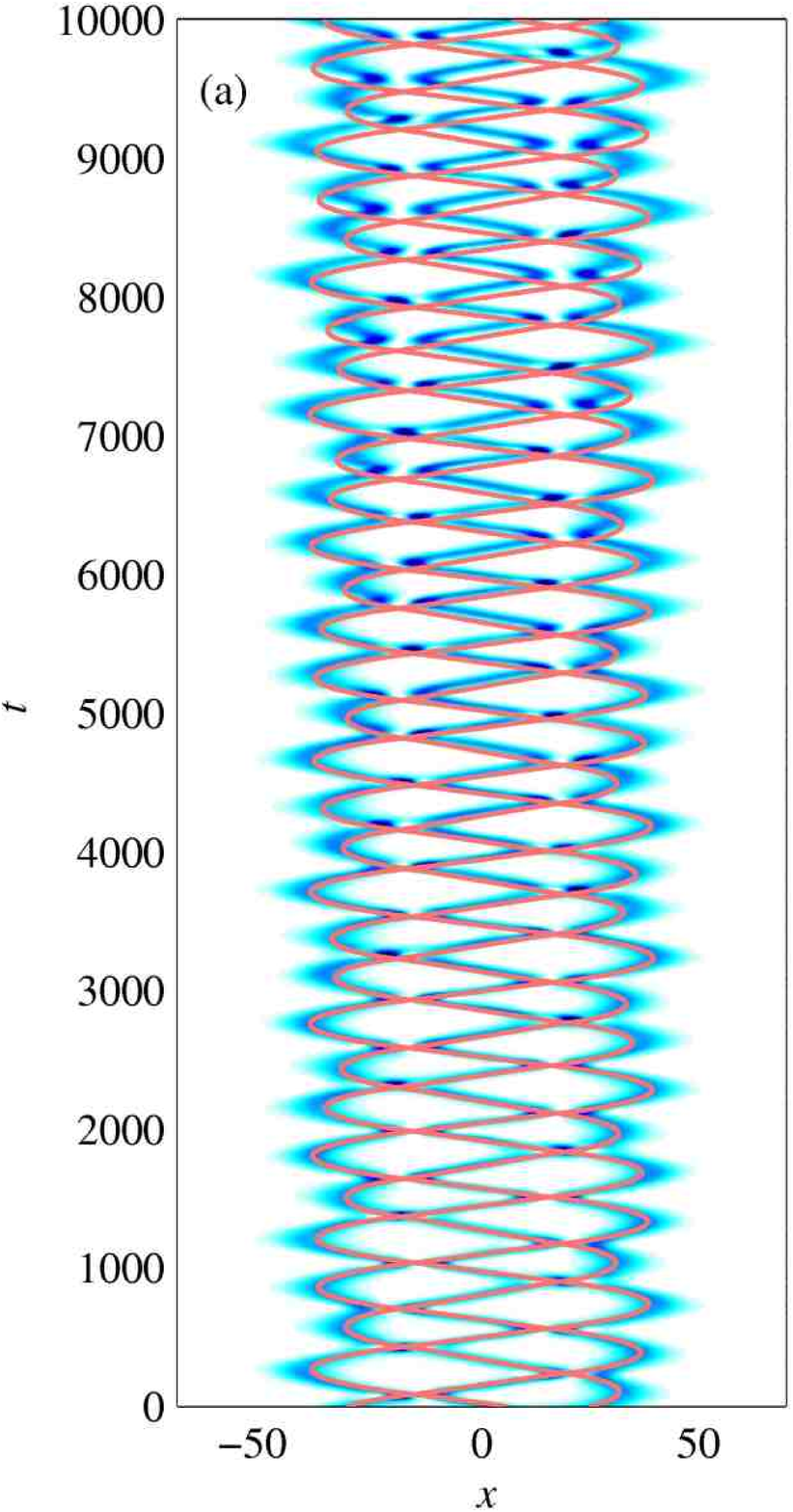}
\includegraphics[height=11.5cm]{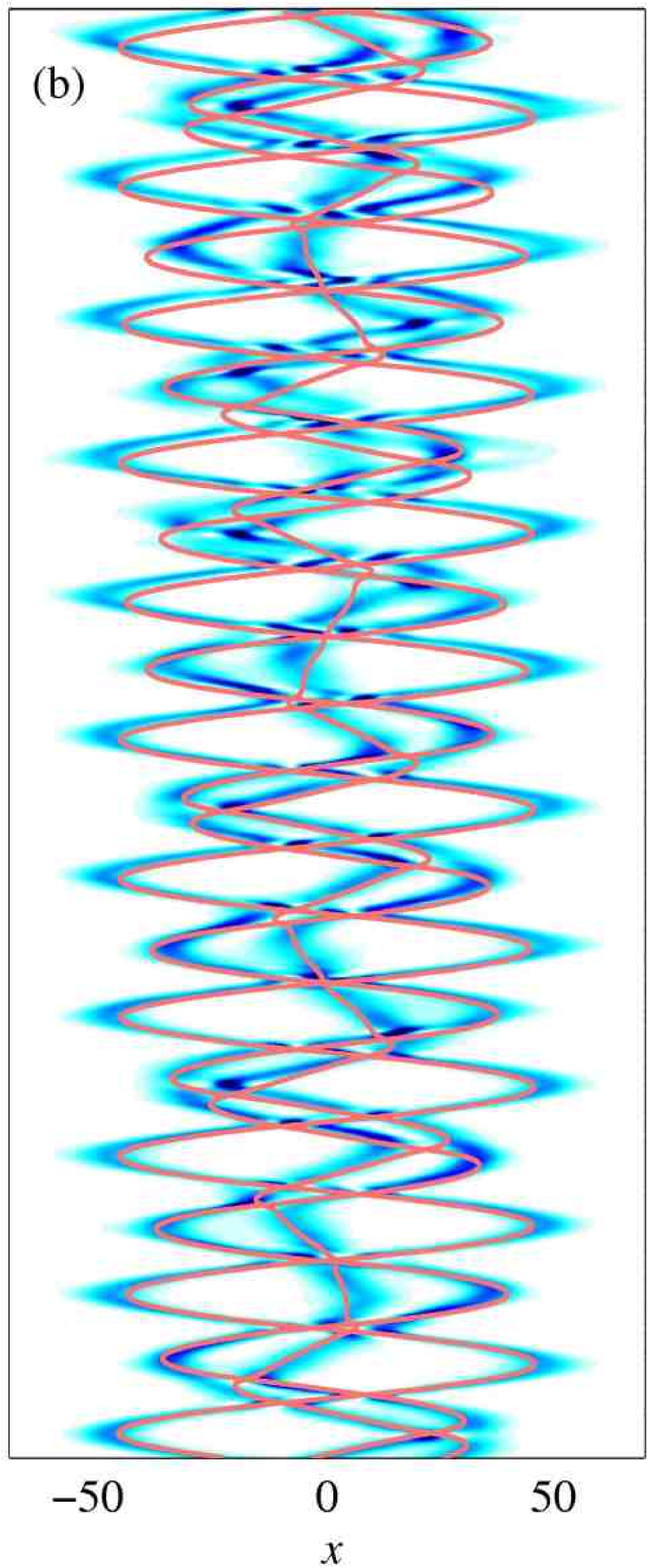}
\includegraphics[height=11.5cm]{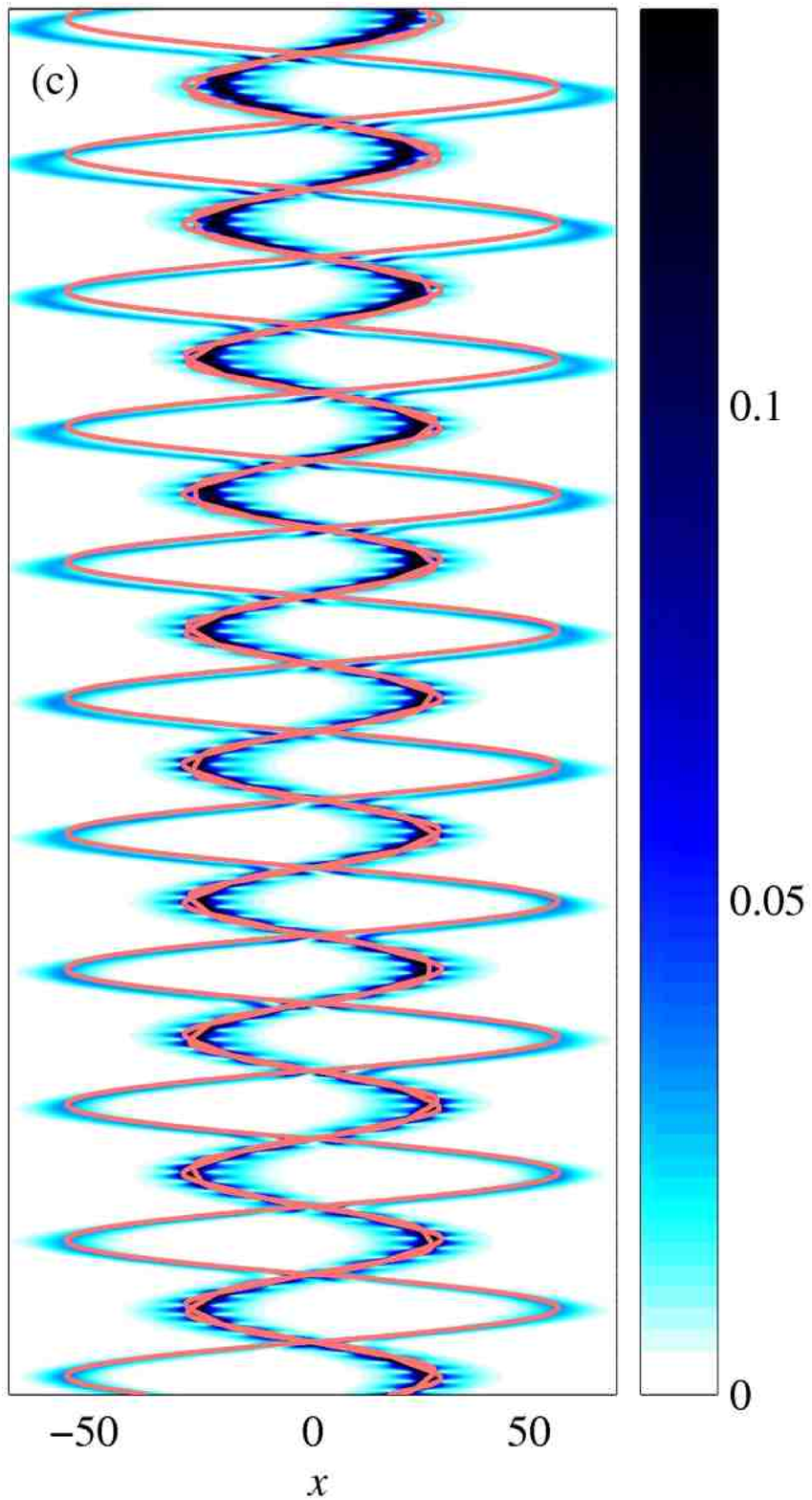}
\caption{(colour online). Trajectories in the particle model (lines) plotted over density distributions predicted by 1D GPE dynamics, corresponding to a regular orbit. The parameters of the system are $\tilde{H}$=10, the solitons have equal effective masses, the axial trapping frequency is $10/2\pi$ Hz, and other parameters (radial trap frequency of $800/2\pi$ Hz, atomic mass and scattering length of $^{7}$Li, and 5000 particles per soliton) correspond to the recent experiment \cite{Strecker_Nature_2002}. The unit of $x$ is then equal to 2.4 $\mu$m, and a unit of $t$ to 0.6 ms. } \label{fig_traj}
\end{figure*}

Figure \ref{fig_traj} shows a comparison of trajectories in the particle model with results from integrations of the 1D GPE [Eq. (\ref{S2})] for the three-soliton system, where the solitons all have equal effective masses. As with the two-soliton case (Sec.\ \ref{Sec:TwoSolitons}), the trajectories in the particle model gradually acquire a shift with respect to the trajectories traced out by the GPE wavefunction peaks.  In Figs.\ \ref{fig_traj}(a) and \ref{fig_traj}(b) the overall shift indicates that subsequent collisions tend to take place sooner in the particle model than is predicted by the GPE evolution; interestingly, in Fig.\ \ref{fig_traj}(c) we observe the opposite, however.   As before, these shifts are caused by an accumulation of small errors, due to the fact that within a harmonic confining potential the individual solitons do not move asymptotically far from each other subsequent to collisions.  In Fig.\ \ref{fig_traj}(c) there is the added complication that two of the solitons appear to have formed a ``bound state.'' 

The comparisons illustrate the good agreement between the particle model and the 1D GPE in the regimes in which the particle model is valid, i.e., when solitons are well separated between collisions [Fig. \ref{fig_traj}(a) and \ref{fig_traj}(b)], even when the motion is chaotic [Fig. \ref{fig_traj}(b)]. When two of the solitons are not well separated [Fig. \ref{fig_traj}(c)], the 1D GPE simulation shows that a ``bound state'' is formed, which looks like a single ``higher-order'' soliton with an excited breathing mode \cite{Panoiu_PRE_1999}. The particle model does not predict well the behaviour within the ``bound state'', but does give a good prediction of the centre-of-mass motion of the ``bound state'' and its interactions with the other soliton; it is likely that the behaviour of the density of the ``bound state'' is strongly coupled to the phase behaviour within the ``bound state.''   As in the two soliton case (Sec.\ \ref{Sec:TwoSolitons}), errors gradually accumulate in the particle model which lead to an overall ``time shift'' in the overall collision dynamics.  It should be noted, however, that apart from this shift qualitative agreement with the dynamics predicted by the GPE remains quite good right up until limits of our numerical calculations ($t=10000$ corresponding to 6 seconds for the experimental parameters in \cite{Strecker_Nature_2002}).

\section{Conclusions}
In this paper we have investigated the soliton-like nature of two and three bright solitary waves in harmonically trapped 1D Bose-Einstein condensates. To this end we employed a model treating each soliton as a classical particle and compared the particle trajectories with simulations of the Gross-Pitaevskii equation. The results from the particle model and the GPE display good agreement when the solitons collide with large relative velocities, such that the collision time is small with respect to the trap period. When the solitons' relative velocities are reduced, the trajectories in the particle model ``get ahead'' of those in the GPE simulations, i.e., points on the particle trajectories can be identified with corresponding points in time during the GPE evolutions, but the GPE evolution is increasingly delayed by comparison; this effect is more pronounced when the solitons collide out of phase, and is due to the non-zero time in which the particle trajectories asymptote to the wave trajectories after collisions. 

We have shown that, when the external potential is harmonic, for systems of two solitons of equal size, the phase difference between the solitons is preserved for repeated collisions. In these systems, repeated out-of-phase collisions can cause the discrepancy between the dynamics in the particle model and the GPE to build up relatively rapidly. We note that it is in this regime, i.e., solitons colliding $\pi$ out of phase,  that the solitons are predicted to be stable against collapse when 3D effects are considered.

Finally, we have extended the treatment to systems of three harmonically trapped solitons. Using the particle model we have shown that, unlike the two soliton systems, systems of three solitons can display chaotic dynamics. Chaotic regimes rescind when the energy of the particle system (with the centre-of-mass motion neglected) is very positive or very negative. In these limits, the dynamics decouple to those of two separated simple harmonic oscillators. The agreement between the dynamics in the particle model 
and the GPE simulations is good in both regular and chaotic regimes.

\section*{ACKNOWLEDGEMENTS}
We thank  J. Brand, S. L. Cornish, K.-P. Marzlin, T. S. Monteiro, N. G. Parker and N. R. Walet for useful discussions, and acknowledge support from the UK EPSRC.

\appendix

\section{From 3D to 1D Gross-Pitaevskii equation \label{App:3Dto1D}}
The one-dimensional GPE can be obtained in approximation from the
three-dimensional GPE by assuming a gaussian ansatz for the radial
wavefunction and integrating out the radial dimensions. The
gaussian ansatz can be taken to be the approximate radial solution
if the radial potential is sufficiently tight that the harmonic
potential energy dominates over the interaction energy in the
radial directions \cite{Salasnich_PRA_2002,Kamchatnov_PRA_2004}.  

Beginning with the 3D GPE:
\begin{equation}
\begin{split}
i\hbar \frac{\partial}{\partial t}\Psi(\mathbf{r})
=&
\left[
-\frac{\hbar^{2}}{2m}\nabla^{2}+V_{\mbox{\scriptsize ext}}(\mathbf{r})+g_{\mbox{\scriptsize 3D}}N|\Psi(\mathbf{r})|^{2}
\right]
\Psi(\mathbf{r})
\label{GPE1}
\end{split}
\end{equation}
where 
\begin{equation}
V_{\mbox{\scriptsize
ext}}(\mathbf{r})=\frac{m}{2}\left[\omega_{x}^{2}x^{2}+\omega_{r}^{2}(y^{2}+z^{2})\right],
\end{equation} and $\Psi(\mathbf{r})$ must have unit norm.
We employ the ansatz $\Psi(\mathbf{r})=\psi(x)\Phi(y)\Phi(z)$, where
\begin{equation}
\Phi(\zeta)=\left(\frac{1}{\sigma^{2}\pi}
\right)^{1/4}\exp\left(\frac{-\zeta^{2}}{2\sigma^{2}}\right)
\end{equation} is the harmonic oscillator ground state, with
$\sigma^{2}=\hbar/m\omega_{r}$. Averaging over the radial degrees of freedom
$\int^{\infty}_{-\infty} dy
dz\Phi^{*}(y)\Phi^{*}(z)$, Eq.\ \ref{GPE1} becomes:
\begin{equation}
i\hbar \frac{\partial}{\partial
t}\psi(x)=
\left[
-\frac{\hbar^{2}}{2m}\frac{\partial^{2}}{\partial x^{2}}
+\frac{m\omega_{x}^{2}x^{2}}{2}
+g_{\mbox{\scriptsize 1D}}N
|\psi(x)|^{2}
+\hbar\omega_{r}
\right]
\psi(x),
 \label{GPE2}
\end{equation} where
\begin{equation}
g_{\mbox{\scriptsize 1D}}=g_{\mbox{\scriptsize 3D}}\int^{\infty}_{-\infty} dydz|\Phi(y)\Phi(z)|^{4}
=2\hbar\omega_{r}a.
\end{equation}

Reassigning the zero of energy allows us to drop the constant term $\hbar\omega_{r}$ from the Hamiltonian [Eq.\ (\ref{GPE2})].  Introducing the dimensionless variables
\begin{equation}
\tilde{x}:=\frac{m|g_{\mbox{\scriptsize 1D}}|N}{\hbar^{2}}x
\end{equation}
and
\begin{equation}
\tilde{t}:=\frac{m|g_{\mbox{\scriptsize 1D}}|^{2}N^{2}}{\hbar^{3}}t,
\end{equation}
Eq.\ (\ref{GPE2}) then becomes:
\begin{equation}
\frac{\partial}{\partial
\tilde{t}}\tilde{\psi}(\tilde{x})=\frac{1}{2}\frac{\partial^{2}}{\partial
\tilde{x}^{2}}\tilde{\psi}(\tilde{x})+\frac{\tilde{\omega}^{2}}{2}\tilde{x}^{2}\tilde{\psi}(\tilde{x})-
|\tilde{\psi}(\tilde{x})|^{2}\tilde{\psi}(\tilde{x})\label{GPE1D_2}.
\end{equation}
We have assumed the $s$-wave scattering length $a$ in $g_{\mbox{\scriptsize 1D}}$ to be negative, and set
\begin{equation}
\tilde{\omega}:=\frac{\omega_{x}\hbar^{3}}{m |g_{\mbox{\scriptsize 1D}}|^{2} N^{2}}
\end{equation}
and
\begin{equation}
\tilde{\psi}(\tilde{x}):=\frac{\hbar}{\sqrt{mgN}}\Psi(x),
\end{equation} 
such that $\tilde{\psi}(\tilde{x})$ is normalised to one with respect to $\tilde{x}$.

In the case where $\tilde{\omega} = 0$, Eq.\ (\ref{GPE1D_2}) has the general soliton solution which takes the form, in the limit $\tilde{t} \rightarrow \pm \infty$ \cite{Faddeev_Book_1987}, 
\begin{equation}
\begin{split}
\tilde{\psi}(\tilde{x})=& \sum_{j}\bigl\{2\eta_{j} \mathrm{sech} \left[2\eta_{j}(\tilde{x}-v_{j}\tilde{t}-x_{0j})\right]\bigr.
\\&
\times \bigl.\exp\left[i v_{j}(\tilde{x}-v_{j}\tilde{t}-x_{0j})+(2\eta_{j}^{2}+v_{j}^{2}/2) \tilde{t}+i\alpha_{0j}\right]\bigr\}
\label{Soliton}
\end{split}
\end{equation}
The normalisation requirement assumed in Eq.\ (\ref{GPE1D_2}) implies that $\sum_{j}\eta_{j}=1/4$.

Throughout the body of the paper, whilst the 1D system is being discussed, the tildes have been dropped for notational convenience.

\section{Harmonically oscillating solution of the 1D GPE \label{App:HarmonicallyOscillating}}

Here we show that an arbitrary solution to the 1D GPE with harmonic potential
\begin{equation}
\psi(x,t)=u(x,t)\exp\left[i\phi(x,t)\right]
\end{equation} can be converted to an oscillating solution with a modified phase: 
\begin{equation}
\psi(x,t)=u(x-\langle x \rangle,t)\exp\left\{i\left[\phi(x,t)+p(t)x-S(t)\right]\right\}.
\end{equation}
This general result is used twice in this paper; firstly to infer the phase behaviour of a single soliton. Secondly it is used to show that any trajectory of two identical solitons is identical, in its collisional density, to the corresponding symmetrical case in which the solitons repeatedly collide in the centre of the potential.  We note in passing that this result holds for any nonlinear Schr\"odinger equation with the nonlinearity given by some function of $|\psi(x)|^{2}$ only.

If $\psi(x,t)=u(x,t)\exp\left[i\phi(x,t)\right]$ is a particular solution to the 1D GPE, where $u(x,t)$ and $\phi(x,t)$ are real-valued functions, the following coupled equations describe their behaviour:
\begin{equation}
\frac{\partial u}{\partial t}=-\frac{1}{2}\left[2\frac{\partial u}{\partial x}\frac{\partial \phi}{\partial x} +\frac{\partial^{2} u}{\partial x^{2}}u\right] \label{WP_1}
\end{equation}
\begin{equation}
-\frac{\partial \phi}{\partial t}=\frac{1}{2}\left(\frac{\partial \phi}{\partial x} \right)^{2}+\frac{1}{u}H(x)u \label{WP_2}
\end{equation}
Defining the functions $\bar{u}(x,t):=u(x+\langle x \rangle,t)$ and $\theta(x,t):=\phi(x + \langle x \rangle)$, where $\langle x \rangle=x_{0}\cos(\omega t)+(p_{0}/\omega)\sin(\omega t)$, and defining the coordinate $\xi=x-\langle x \rangle$ the following relations follow trivially:
\begin{equation}
\bar{u}(\xi,t)=u(x,t),
\end{equation}
\begin{equation}
\theta(\xi,t)=\phi(x,t),
\end{equation} 
\begin{equation}
\frac{\partial u(x,t)}{\partial x}=\frac{\partial \bar{u}(\xi,t)}{\partial\xi},
\end{equation}
\begin{equation}
\frac{\partial u(x,t)}{\partial t}=-\dot{\langle x \rangle}\frac{\partial \bar{u}(\xi,t)}{\partial\xi} +\frac{\partial \bar{u}(\xi,t)}{\partial t}.
\end{equation}  Also, the relationship between the partial derivatives of $\theta(\xi,t)$ and $\phi(x,t)$ is the same as that between $\bar u(\xi,t)$ and $u(x,t)$. From now on it is assumed that $\bar{u}$, $\theta$ and their derivatives are be functions of $\xi$ and $t$.

Eqs.\ (\ref{WP_1}) and (\ref{WP_2}) become:
\begin{equation}
\frac{\partial \bar{u}}{\partial t}=-\frac{1}{2}\left[2\frac{\partial\bar{u}}{\partial t}\left(\frac{\partial \theta}{\partial \xi}-\dot{ \langle x \rangle}\right)+\frac{\partial^{2}\theta}{\partial \xi^{2}}\bar{u} \right]
\end{equation} 
and
\begin{equation}
\frac{\partial \theta}{\partial \xi}\dot{\langle x \rangle}-\frac{\partial\theta}{\partial t}=\frac{1}{2}\left(\frac{\partial \theta}{\partial \xi} \right)^{2}+\frac{1}{\bar{u}}H(\xi)\bar{u} +\frac{\omega^{2}}{2}\langle x \rangle^{2}+\omega^{2}\xi\langle x \rangle.
\end{equation} 
By choosing $\bar{\phi}(\xi,t)$ such that 
\begin{equation}
\frac{\partial\bar{\phi}}{\partial\xi}=\frac{\partial\theta}{\partial\xi}-\dot{\langle x\rangle},
\end{equation}
and
\begin{equation}
\frac{\partial\bar{\phi}}{\partial t}=\frac{\partial\theta}{\partial t}-\xi \ddot{\langle x\rangle}+\frac{\partial S(t)}{\partial t},
\end{equation} where 
\begin{equation}
S(t)=\left( p_{0}^{2}-\frac{x_{0}^{2}\omega^{2}}{2}\right)\frac{\sin(2\omega t)}{2\omega}+\frac{x_{0}p_{0}}{2}\cos(2\omega t)-\frac{x_{0}p_{0}}{2} + S(0),
\end{equation}
we can relabel $\xi$ as $x$, and find that $\bar{\phi}(x,t)$ and $\bar{u}(x,t)$ are solutions of Eqs.\ (\ref{WP_1}) and (\ref{WP_2}). Hence, $\bar{u}(x,t)$ has the profile of $u(x,t)$, but undergoes additional global harmonic oscillations at the trap frequency, and the result is proved.

\section{Motion of a single trapped soliton \label{App:Motion}}
The motion of a single soliton can be derived by adapting the method of Scharf and Bishop \cite{Scharf_PRA_1992} where a solution to the homogeneous GPE is used as an ansatz for the GPE with an external potential. This method has the advantage that it can be used with other than harmonic  potentials. Here we show that for a harmonic potential it confirms the expected result of simple harmonic motion.

In the 1D GPE, the norm $\mathcal{N}$, and energy $\mathcal{E}$, are conserved quantities. The single soliton solution of the homogeneous case [Eq.\ (\ref{Sol_1}) with $j=1$] is used as an ansatz for the harmonic case.   Evaluating the norm and energy functionals  with this ansatz yields:
\begin{equation}
\mathcal{N}=\int^{\infty}_{-\infty} dx |\psi(x)|^{2}=4\eta
\end{equation}
\begin{equation}
\begin{split}
\mathcal{E}=&\int^{\infty}_{-\infty}dx 
\left[
\frac{1}{2}\left|\frac{\partial\psi(x)}{\partial x}\right|^{2}+|\psi(x)|^{2}\frac{\omega^{2}x^{2}}{2}-\frac{1}{2}|\psi(x)|^{4}
\right]
\\=&2\eta\dot{q}^{2}-\frac{16}{5}\eta^{3}+\frac{\omega^{2}}{2}\left(4\eta q^{2}+\frac{\pi^{2}}{12\eta}\right).
\end{split}
\end{equation}
The conservation of the norm $\mathcal{N}$, leads to
$\eta=$constant,
and the conservation of the energy $\mathcal{E}$, leads to an equation of motion for the peak of the soliton, $q$:
\begin{equation}
\ddot{q}=-\omega^{2} q. \label{N1}
\end{equation}

\section{Deduction of the inter-particle potential \label{App:Deduction}}
In order to deduce an effective inter-particle potential, it is necessary to perform a classical inverse-scattering calculation to find the potential which produces the same asymptotic position-shifts as given in Ref.\ \cite{Zakharov_ZETF_1971} for solitons emerging from a collision. This proof follows the method given in Ref.\ \cite{Maki_PRL_1986}. 

The position-shifts given in Ref.\ \cite{Zakharov_ZETF_1971} are equivalent to the asymptotic time-shifts for two solitons of initial speeds $v_{1}$ and $v_{2}$, and effective masses $\eta_{1}$ and $\eta_{2}$, given by
\begin{equation}
\Delta t=-\frac{1}{2\left(v_{1}-v_{2}\right)}\left( \frac{1}{\eta_{1}}+\frac{1}{\eta_{2}} \right) \ln \left[\frac{\left(v_{1}-v_{2}\right)^{2}+4\left(\eta_{1}+\eta_{2}\right)^{2}}{\left(v_{1}-v_{2}\right)^{2}+4\left(\eta_{1}-\eta_{2}\right)^{2}} \right]. 
\label{Time_wave}
\end{equation}

We wish to produce these shifts in a system of two classical particles described by the Hamiltonian:
\begin{equation}
H:=\frac{p^{2}}{2\mu_{r}} +V(q) \label{H_scatter}
\end{equation} where $q:=q_{1}-q_{2}$ is the relative coordinate, $\mu_{r}:=\eta_{1}\eta_{2}/(\eta_{1}+\eta_{2})$ is the reduced effective mass, $p=(\eta_{1}p_{2}-\eta_{2}p_{1})/(\eta_{1}+\eta_{2})$ is the relative momentum, and the centre of mass has been separated from the problem. 
For particles initially separated at infinity,  and noting that $p^{2}/2\mu_{r}=\mu_{r}\dot{q}^{2}/2$, this Hamiltonian takes the asymptotic form 
\begin{equation}
H=\frac{\mu_{r}}{2}\left(v_{1}-v_{2}\right)^{2}:=E_{\infty},
\end{equation} 
i.e., we assume the potential must vanish asymptotically.

By rearranging Eq.\ (\ref{H_scatter}), we may write the infinitesimal
\begin{equation}
dq=\sqrt{\frac{2\left[E_{\infty}-V(q)\right]}{\mu}} dt,
\end{equation} 
since energy is conserved over the whole trajectory.  Integrating the time difference between trajectories with and without the inter-particle potential, we determine the asymptotic timeshift to be
\begin{equation}
\Delta t=\left(\frac{\mu}{2}\right)^{1/2}\int_{-\infty}^{\infty} 
dq
\frac{1}{E_{\infty}^{1/2}}\left[\left(1-\frac{V(q)}{E_{\infty}}\right)^{-1/2}-1 \right]. 
\label{Time_particle}
\end{equation}

Now, expanding Eqs.\ (\ref{Time_wave}) and (\ref{Time_particle}) in terms of powers of $1/E_{\infty}$, and equating equal powers, we obtain
\begin{equation}
\begin{split}
\int_{-\infty}^{\infty}dq V(q)^{n} =&
\frac{1}{2}\left(\frac{1}{\eta_{1}}+\frac{1}{\eta_{2}} \right)
\\&\times
\left\{\left[4\eta_{1}\eta_{2}\left(\eta_{1}+\eta_{2}\right) \right]^{n}-\left[\frac{ 4\eta_{1}\eta_{2}\left(\eta_{1}-\eta_{2} \right)^{2}}{\eta_{1}+\eta_{2}} \right] ^{n}\right\}\\
&\times\frac{(-1)^{n}2^{n-1}\left[(n-1)!\right]^{2}}{\left(2n-1 \right)!}
\end{split}\label{V_int1}
\end{equation} for all positive integers $n$.  We now evaluate the integral of a candidate potential:
\begin{equation}
\begin{split}
\int_{-\infty}^{\infty}dqV(q)^{n} =&\int_{-\infty}^{\infty}dq \left[-2(\eta_{1}+\eta_{2})\eta_{1}\eta_{2}\mathrm{sech}^{2}\left(\frac{2\eta_{1}\eta_{2}}{\eta_{1}+\eta_{2}}q\right)\right]^{n}\\
=&
\frac{1}{2}\left(\frac{1}{\eta_{1}}+\frac{1}{\eta_{2}} \right)\left[4\eta_{1}\eta_{2}\left(\eta_{1}+\eta_{2}\right) \right]^{n} 
\\&\times
\frac{(-1)^{n}2^{n-1}\left[(n-1)!\right]^{2}}{\left(2n-1 \right)!}
\end{split}\label{V_int2}
\end{equation} 
Comparing the expressions (\ref{V_int1}) and (\ref{V_int2}) it follows that the potential 
\begin{equation}V(q)=-2(\eta_{1}+\eta_{2})\eta_{1}\eta_{2}\mathrm{sech}^{2}\left(\frac{2\eta_{1}\eta_{2}}{\eta_{1}+\eta_{2}}q\right)\end{equation} gives the correct time shift in the limit $2|\eta_{1}-\eta_{2}|\ll |v_{1}-v_{2}|$.

\section{Collisional form preservation with two harmonically trapped solitons \label{App:CollisionalForm}}

Let us consider a general parity operator $\hat{P}(\phi)$, such that $\hat{P}(\phi)\psi(x,t)=e^{i\phi}\psi(-x,t):=\chi(x,t)$. We want to know whether, if $\chi(x,t_{0}) = \psi(x,t_{0})$, it continues to be the case that $\chi(x,t) = \psi(x,t)$ for all $t$.  

We take the time derivative of $\chi(x,t)$. Noting that $\hat{P}(\phi)|\psi(x,t)|^{2}\psi(x,t)=e^{i\phi}|\psi(-x,t)|^{2}\psi(-x,t)=|\hat{P}(\phi)\psi(x,t)|^{2}[\hat{P}(\phi)\psi(x,t)]$ we deduce from
the 1D Gross-Pitaevskii equation [Eq.\  (\ref{S2})] that
\begin{equation}
\begin{split}
i\frac{\partial}{\partial t}\chi(x,t) = &
i\hat{P}(\phi)\frac{\partial}{\partial t}\psi(x,t) 
\\=& 
\hat{P}(\phi)
\left\{
\left[
-\frac{1}{2}\frac{\partial^{2}}{\partial x^{2}}
+\frac{\omega^{2}x^{2}}{2}
-|\psi(x,t)|^{2}
\right]\psi(x,t)
\right\}
\\
= &
\left[
-\frac{1}{2}\frac{\partial^{2}}{\partial x^{2}}
+\frac{\omega^{2}x^{2}}{2}
-|\chi(x,t)|^{2}
\right]\chi(x,t).
\end{split}
\end{equation}
Hence, we see that the time-evolutions of $\psi(x,t)$ and $\hat{P}(\phi)\psi(x,t)=\chi(x,t)$ are governed by the same differential equation.  If we also choose an initial condition such that $\psi(x,t_{0})=\chi(x,t_{0})$, it must therefore follow that $\chi(x,t)=\psi(x,t)$ for all $t$. In other words, parity is conserved in the sense that an initially symmetric wave function will have that symmetry preserved throughout its subsequent dynamical evolution.

In this paper, the most important consequence of this result is that a system of two identical solitons with equal and opposite velocities will repeatedly collide with the exact same collisional form (e.g., in phase or $\pi$ out of phase) at the exact centre of the trapping potential. Using the results of appendix \ref{App:HarmonicallyOscillating}, it follows that an equivalent result holds upon the addition of a centre of mass oscillation.

\end{document}